\documentclass[aps,prd,reprint,nofootinbib,superscriptaddress,longbibliography,floatfix]{revtex4-2}
\usepackage[T1]{fontenc}
\usepackage{lmodern,microtype,amsmath,amssymb,mathtools,bm,mathrsfs}
\usepackage{booktabs,array,graphicx,tikz}
\usetikzlibrary{decorations.pathmorphing,decorations.markings,arrows.meta}
\usepackage{xcolor}
\definecolor{linkblue}{RGB}{0,90,210}
\definecolor{citegreen}{RGB}{0,145,100}
\usepackage[colorlinks=true,linkcolor=linkblue,citecolor=citegreen,urlcolor=linkblue]{hyperref}
\hypersetup{pdftitle={Radiative dynamics of compact binaries in massless scalar–tensor theories via effective field theory methods},pdfauthor={Gabriel Luz Almeida and Mehbub Rehman Khan}}
\newcommand{\dd}{\mathrm d}\newcommand{\ii}{\mathrm i}
\newcommand{\STF}{\operatorname{STF}}
\newcommand{\vk}{\mathbf k}\newcommand{\vq}{\mathbf q}\renewcommand{\vr}{\mathbf r}
\newcommand{\vx}{\mathbf x}\newcommand{\vv}{\mathbf v}\newcommand{\va}{\mathbf a}\newcommand{\vn}{\mathbf n}

\newcommand{\order}{\mathcal O}
\allowdisplaybreaks[2]
\makeatletter
\AtBeginDocument{\immediate\write\@auxout{\string\citation{apsrev42Control}}}
\makeatother
\begin{document}
\title{Radiative dynamics of compact binaries in massless scalar–tensor theories \\ via effective field theory methods}
\author{Gabriel Luz Almeida}
\email{galmeida@seu.edu.cn}
\affiliation{School of Physics {\itshape\&} Shing-Tung Yau Center, Southeast University, Nanjing 210096, China}
\author{Mehbub Rehman Khan}
\email{mrk104@pitt.edu}
\affiliation{Department of Physics and Astronomy, University of Pittsburgh, Pittsburgh, PA 15260, USA}

\begin{abstract}
We derive the tensor and scalar source multipole moments
governing radiation from nonspinning compact binaries in
massless scalar--tensor theories using effective field
theory (EFT). Retaining generic scalar charges and
nonlinear compact-body responses, we compute the
instantaneous tensor and scalar energy fluxes through
second post-Newtonian order relative to their leading
quadrupole and dipole contributions, respectively.
We give explicit results for generic orbits and their
circular limit.
The calculation includes an EFT derivation of the leading
scalar dipole radiation-reaction force and its associated
center-of-mass correction. For the tensor flux, we
incorporate the nonlinear relation between source and
radiative multipoles from the literature.
Our scalar flux independently confirms established
instantaneous results through the retained orders,
while the general relativity limit reproduces the known
instantaneous flux through second post-Newtonian order
for generic orbits.
The explicit matching presented here establishes
a systematic EFT framework for extending these results
to higher radiative orders in scalar--tensor theories.
\end{abstract}

\maketitle
\section{Introduction}\label{sec:introduction}

Compact binaries probe both the propagation of gravity and
the response of strongly self-gravitating bodies to their
environment. In massless scalar--tensor theories, an additional
long-range field modifies the conservative interaction and
opens a scalar radiation channel. Because the scalar charge
need not be proportional to the inertial mass, a binary can
emit dipole radiation even though its leading tensor radiation
is quadrupolar. This connection between internal structure,
orbital dynamics, and radiation makes scalar--tensor theories
a particularly instructive setting for precision tests of
gravity.

The formulation of scalar--tensor gravity developed from the
work of Fierz, Brans and Dicke, Bergmann, and
Wagoner~\cite{Fierz1956,BransDicke1961,Bergmann1968,Wagoner1970}.
For compact objects, an effective description replaces each
extended body by a point particle with a scalar-dependent
mass~\cite{Eardley1975,DEF1992}, whose derivatives determine
the scalar charge and higher response coefficients.
Strong self-gravity can amplify these responses, as occurs
in spontaneous
scalarization~\cite{DEF1993,DEF1996,Scalarization2024}.
During binary evolution, dynamical scalarization further
illustrates why a weak asymptotic coupling need not imply
small strong-field effects~\cite{Barausse2013,Palenzuela2014}.
Here, we adopt the adiabatic point-particle
expansion about a fixed asymptotic scalar value.

These aspects of scalar--tensor gravity are constrained by
observations in complementary regimes. Solar-system
measurements test weak-field deviations, while binary
pulsars probe scalar radiation and strong-field
responses~\cite{Cassini2003,WillZaglauer1989,Freire2012,Antoniadis2013,Kramer2021}.
Gravitational-wave observations extend these tests to the
relativistic inspiral~\cite{LIGO2016,TestsGWTC3,Berti2015}.
Together, they constrain different combinations of theory
parameters and body-dependent responses. Cosmological
evolution also plays a role, since it can drive certain
scalar--tensor models toward the general-relativistic
regime~\cite{DamourNordtvedt1993}. The radiative signatures
depend on the responses of both bodies: equal scalar
charges suppress the leading dipole without generally
eliminating scalar quadrupole radiation.

The post-Newtonian (PN) description of these systems has
been developed through several complementary methods.
Mirshekari and Will obtained the equations of motion
through 2.5PN, including the leading dipolar radiation
reaction at 1.5PN~\cite{MirshekariWill2013}, while Lang
computed tensor wave generation through 2PN and scalar
waves and energy flux at the corresponding
orders~\cite{Lang2014,Lang2015}.
Building on the description
of binary radiation, Sennett, Marsat, and Buonanno organized
the waveform and phase for quasicircular
inspirals~\cite{Sennett2016}. The multipolar
post-Minkowskian treatment of Bernard, Blanchet, and Trestini
includes wave-generation terms through 1.5PN beyond the
usual quadrupole, together with hereditary
contributions~\cite{BBT2022}. More recent work has extended
the quasi-Keplerian description, orbit-averaged radiation
from eccentric binaries, and angular-momentum
flux~\cite{Trestini2024,Trestini2025,JainRettegno2025}.
These results provide benchmarks for an independent
effective field theory (EFT) construction of the sources
and their matching. Such a construction also identifies the 
interactions required to extend the calculation to
higher orders.

The conservative sector has seen parallel advances.
The equations and integrals of motion are known through
3PN~\cite{Bernard2018,Bernard2019}, with an
effective-one-body description developed in
Ref.~\cite{Julie2023}. Within EFT, the conservative
dynamics has been derived through 2PN~\cite{Almeida2024},
while the static potential has been extended to 5PN
using multiloop techniques~\cite{Almeida2026static}
and recently to 9PN for general scalar-dependent
masses~\cite{AlmeidaLiuYuan9PN} using the position-space
framework developed for general relativity~\cite{Liu:2026sib}.

Alongside these advances in post-Newtonian accuracy, the
EFT framework has been extended to scalar--curvature
couplings and finite-size effects. In
Einstein--scalar--Gauss--Bonnet gravity, calculations
include spin-dependent conservative effects through
3PN~\cite{AlmeidaZhou2024} and conservative scattering
through third post-Minkowskian (3PM)
order~\cite{AlmeidaDuLiuWang2026}. By retaining the full
velocity dependence at each order in Newton's constant,
the PM expansion complements the PN description of bound
motion. Finite-size effects have also been investigated:
Bernard's calculation of dipolar tides~\cite{Bernard2020}
has been followed by higher-order conservative and
radiative tidal
results~\cite{BernardDonesMougiakakos2024,DonesBernard2026}
and by a 3PM treatment including scalar--Gauss--Bonnet
interactions~\cite{LiuWang2026tidal}. These developments
motivate a common EFT organization of orbital dynamics
and radiation. Here we focus on the nonspinning,
point-particle sector with a minimally coupled massless
scalar; curvature couplings and tidal operators require
additional interactions.

The organizing principle of this EFT description is the
hierarchy between the size of a compact object, the binary
separation, and the radiation
wavelength~\cite{GoldbergerRothstein2006,Goldberger2007,FoffaSturani2014,Porto2016,Levi2020}.
Potential modes determine the orbital dynamics, while
radiation modes couple to composite multipole operators.
The conservative 2PN calculation~\cite{GilmoreRoss2008}
and the nonrelativistic field
parametrization~\cite{KolSmolkin2008} illustrate the
efficiency of this organization. Its application to
radiation was developed through radiative corrections
and the multipole expansion at the level of the
action~\cite{GoldbergerRoss2010,Ross2012}.
Kuntz, Piazza, and Vernizzi extended the framework to
scalar--tensor radiation~\cite{Kuntz2019}. In general
relativity, Leibovich \emph{et al.} obtained the 2PN
radiation sources and flux in a direct metric-field
calculation~\cite{Leibovich2020}. The general-dimensional
multipole action~\cite{Amalberti2024d} and the subsequent
3PN radiation calculation~\cite{Amalberti2024flux}
provide a framework for dimensional regularization
and higher-order matching.

Predicting the observed inspiral requires the radiative
description to accompany the conservative dynamics:
energy loss and wave amplitudes must be controlled at
compatible orders. Although the results reviewed above
establish important PN information, the systematic
construction and matching of radiation sources within
scalar--tensor EFT remain less developed than the
conservative sector. We advance this construction by
deriving the local radiation couplings in a common gauge
and with a common regularization prescription.
We organize the diagrams by topology, evaluate their
contributions to the gravitational pseudotensor and
scalar source, and construct the source multipole moments.

The calculation retains tensor radiation through
quadrupole-relative 2PN and scalar radiation through
dipole-relative 2PN. Since scalar dipole emission begins
one PN order earlier, the latter corresponds to
quadrupole-relative 1PN. We present the moments in the
linear harmonic gauge before and after imposing the
center-of-mass condition, together with the
instantaneous power for generic orbits and its
circular limit. To evaluate the multipole derivatives
at these orders, we supplement the conservative
dynamics with the leading dipolar radiation-reaction
force. We derive this force within the causal EFT
framework~\cite{GalleyTiglio2009,GalleyLeibovich2012}
and determine the associated dissipative correction
to the center-of-mass frame.

The tensor flux also receives nonlinear propagation
contributions from the scalar field. These enter
through the relation between source and radiative
multipoles, which we adopt from
Ref.~\cite{BBT2022}. In particular, we include the
scalar-induced memory and its accompanying local terms;
the derivative of the memory integral contributes
to the instantaneous flux. This nonlinear
radiative-moment relation is an input from the
literature, not a result of our EFT matching.
Its independent derivation within EFT lies outside
the scope of this work and is left for future study.
Tail contributions remain separate from the
instantaneous flux considered here.

The scalar instantaneous flux agrees with established
generic-orbit results through the retained
orders~\cite{BBT2022}, providing an independent
confirmation within EFT.
In the tensor sector, combining our source calculation
with the adopted nonlinear radiative-moment relation
reproduces the corresponding local contributions at
relative 1.5PN~\cite{BBT2022}. 
The general relativity limit also reproduces the known
instantaneous flux through 2PN for generic
orbits~\cite{GopakumarIyer1997}.
These comparisons test the source matching, dynamical
reduction, and center-of-mass transformation in the
combinations entering the flux. Beyond these checks,
the explicit sources and multipoles provide the
ingredients for systematic extensions of the radiative
calculation. 
Higher-order radiation reaction and the incorporation
of tail contributions into orbital evolution and phasing
remain future applications of this framework.

The paper is organized as follows.
Section~\ref{sec:setup} introduces the EFT and its
multipole expansion. Section~\ref{sec:calculation}
describes the topology classes and source calculation.
Section~\ref{sec:dynamics} presents the conservative
dynamics, the leading dipolar radiation reaction,
and the center-of-mass transformation.
Section~\ref{sec:results} gives the source multipoles,
and Sec.~\ref{sec:flux} combines them with the required
nonlinear propagation terms to obtain the instantaneous
energy loss. We conclude in Sec.~\ref{sec:conclusions}.
Appendix~\ref{app:action} details the conservative
dynamics, Noether charges, and center-of-mass construction.
Appendix~\ref{app:coefficients} specifies the multipole
and flux reconstruction formulas and derives their
circular reduction. The explicit coefficient expressions
are supplied in the ancillary Mathematica notebook.

\textit{Conventions.}
We use metric signature $(+,-,-,-)$.
Greek indices label spacetime components and Latin
indices label spatial components, with Euclidean
contractions for spatial vectors. The positions,
velocities, and accelerations of the bodies are
$\vx_A(t)$, $\vv_A=\dot{\vx}_A$, and
$\va_A=\dot{\vv}_A$, with analogous definitions for
body $B$. The relative variables are
$\vr=\vx_A-\vx_B$, $r=|\vr|$, $\vn=\vr/r$,
$\vv=\vv_A-\vv_B$, $v^2=\vv\cdot\vv$, and
$\dot r=\vn\cdot\vv$. 
Potential integrals are evaluated using dimensional
regularization, with the number of spatial dimensions
continued to $d=3+\epsilon$ and
$G_d=G\mu_{\rm DR}^{-\epsilon}$, where $\mu_{\rm DR}$
is an inverse-length scale. Physical moments and fluxes
are given at $d=3$. We set $c=1$ unless it is used to
help identify PN orders: a factor $c^{-2n}$
denotes a relative $n$PN correction, with integer
or half-integer $n$ as appropriate.
A bracket $[n]$ labels the corresponding
coefficient, while $(v^n)$ labels an absolute velocity
order in a source. Sums over bodies run over $A$ and $B$;
in ordered-body expressions, $(A\leftrightarrow B)$
exchanges every body label and reverses $\vr$.
\section{Effective field theory}\label{sec:setup}

\subsection{Scales, fields, and compact-body response}

During the slow inspiral, the characteristic body size $r_s$,
orbital separation $r$, and radiation wavelength $\lambda$
satisfy
\begin{equation}
r_s\ll r\ll\lambda,\qquad
\frac{r_s}{r}\sim\frac{v^2}{c^2},\qquad
\frac{r}{\lambda}\sim\frac{v}{c}.
\label{eq:scales}
\end{equation}
This hierarchy provides the organizing principle of the EFT
construction. The ratios between these scales determine the
power counting in $v/c$, with each PN order corresponding
to a factor of $v^2/c^2$. Dissipative effects also introduce
half-integer PN orders, beginning with the leading scalar
dipole reaction at 1.5PN.

Integrating out the
internal scale replaces each compact body by a point particle
whose mass and response coefficients encode its internal
structure. Additional worldline operators account for
finite-size effects at higher orders. At the accuracy
considered here, we retain the nonspinning monopolar action.

In the Einstein frame, this description is specified by
\begin{align}
S&=S_{\rm EH}+S_\varphi+S_{\rm pp},
\label{eq:action}\\
S_{\rm EH}
&=-\frac{2}{\kappa^2}
\int\dd^{d+1}x\sqrt{-g}\,R,
\nonumber\\
S_\varphi
&=\frac{4}{\kappa^2}
\int\dd^{d+1}x\sqrt{-g}\,
g^{\mu\nu}\partial_\mu\varphi\partial_\nu\varphi,
\nonumber\\
S_{\rm pp}
&=-\sum_A\int m_A(\varphi)\,\dd s_A.
\nonumber
\end{align}
Here $R$ is the Ricci scalar, $g$ the determinant of the metric,
$\varphi$ the dimensionless massless scalar, and $\dd s_A$ the
proper interval along body $A$'s worldline. We use the
field-normalization constant $\kappa=\sqrt{32\pi G_d}$.

The remaining hierarchy, $r\ll\lambda$, separates the fields
responsible for the orbital interaction from those describing
radiation. We implement this separation by expanding the
metric about flat spacetime and the scalar about its constant
asymptotic value $\varphi_0$:
\begin{align}
g_{\mu\nu}
&=\eta_{\mu\nu}+\kappa H_{\mu\nu}+\kappa \bar h_{\mu\nu},
\nonumber\\
\varphi
&=\varphi_0+\kappa \Phi+\kappa\bar\varphi.
\label{eq:fields}
\end{align}
Here $\eta_{\mu\nu}$ is the Minkowski metric,
$H_{\mu\nu}$ and $\Phi$ are potential fields, and
$\bar h_{\mu\nu}$ and $\bar\varphi$ are radiation fields.
Their different roles are reflected in their momentum
scalings. Writing $(q^0,\vq)$ for a potential four-momentum
and $(k^0,\vk)$ for a radiation four-momentum, we have
$(q^0,\vq)\sim(v/r,1/r)$ and
$(k^0,\vk)\sim(v/r,v/r)$.
Radiation fields are therefore slowly varying across the
binary and can be Taylor expanded about a reference point
within the source, chosen at the center of mass when
specializing to that frame. 

Integrating out the potential modes determines the orbital
interactions and the multipole couplings of the binary
to radiation. The subsequent treatment of the radiation
fields with retarded boundary conditions also determines
their reaction on the orbital motion~\cite{Almeida:2021xwn,Almeida:2022jrv,Almeida:2026clf}.

The scalar normalization in Eq.~\eqref{eq:fields} leaves the
quadratic kinetic term in the form $4(\partial\Phi)^2$.
When a canonical normalization is needed, we use
$\vartheta=\sqrt{8}\,\Phi$ for the potential field and
$\bar\vartheta=\sqrt{8}\,\bar\varphi$ for the radiation field.

The effect of strong self-gravity on the coupling to the
scalar is encoded in the worldline mass function
$m_A(\varphi)$. Its value at the asymptotic background defines
$m_A\equiv m_A(\varphi_0)$, while its logarithmic derivatives
determine the scalar charge and higher response coefficients:
\begin{align}
\alpha_A
&=\left.\frac{\dd\log m_A}{\dd\varphi}\right|_{\varphi_0},
\qquad
\beta_A
=\left.\frac{\dd\alpha_A}{\dd\varphi}\right|_{\varphi_0},
\nonumber\\
\beta'_A
&=\left.\frac{\dd\beta_A}{\dd\varphi}\right|_{\varphi_0},
\label{eq:sensitivities}
\end{align}
and so on. These coefficients enter the interaction vertices through
the expansion in $\delta\varphi=\varphi-\varphi_0$,
\begin{align}
\frac{m_A(\varphi)}{m_A}
={}&1+\alpha_A\delta\varphi
+\frac12(\alpha_A^2+\beta_A)\delta\varphi^2
\nonumber\\
&+\frac16
(\alpha_A^3+3\alpha_A\beta_A+\beta'_A)
\delta\varphi^3
+\order(\delta\varphi^4).
\label{eq:mass-derivatives}
\end{align}
Because they are evaluated at the fixed background
$\varphi_0$, the response coefficients are treated as constants
under orbital time differentiation. Their leading contribution
to the binary interaction combines with tensor exchange to give
the effective Newtonian coupling
$G_{AB}=G(1+\alpha_A\alpha_B)$.
To express the resulting two-body dynamics, we introduce
the total mass $m=m_A+m_B$, the reduced mass
$\mu=m_Am_B/m$, the symmetric mass ratio $\nu=\mu/m$,
and the relative mass difference $\Delta=(m_A-m_B)/m$.

The higher-order dependence on the scalar responses can be
written compactly using the standard dimensionless
combinations
\begin{align}
\bar\gamma_{AB}
&=-\frac{2\alpha_A\alpha_B}{1+\alpha_A\alpha_B},
&
\bar\beta_A
&=\frac{\beta_A\alpha_B^2}
{2(1+\alpha_A\alpha_B)^2},
\label{eq:ST-parameters1}\\
\delta_A
&=\frac{\alpha_A^2}{(1+\alpha_A\alpha_B)^2},
&
\epsilon_A
&=\frac{\beta'_A\alpha_B^3}
{(1+\alpha_A\alpha_B)^3},
\nonumber\\
\zeta_{AB}
&=\frac{\beta_A\beta_B\alpha_A\alpha_B}
{(1+\alpha_A\alpha_B)^3}.
&&
\label{eq:ST-parameters2}
\end{align}
The corresponding body-$B$ combinations follow by
interchanging the body labels. We use these parameters
in the conservative dynamics and circular-orbit expressions
without assuming small scalar charges.

\subsection{Gauge fixing and radiation action}

To describe radiation at wavelengths much larger than the
binary separation, we match the orbital-scale theory onto
an effective description in which the binary acts as a
localized source. This requires integrating out the potential
modes while keeping the radiation fields fixed. For the metric,
the radiation background is
$\bar g_{\mu\nu}=\eta_{\mu\nu}+\kappa\bar h_{\mu\nu}$.
We perform this integration in background-covariant linear
harmonic gauge, implemented by the gauge-fixing action
\begin{align}
S_{\rm GF}
&=\int\dd^{d+1}x\sqrt{-\bar g}\,
\bar g^{\mu\nu}\Gamma_\mu\Gamma_\nu,
\nonumber\\
\Gamma_\mu
&=\bar\nabla_\alpha H_\mu{}^\alpha
-\frac12\bar\nabla_\mu H_\alpha{}^\alpha.
\label{eq:gauge}
\end{align}
Here $\bar\nabla$ is the covariant derivative associated with
$\bar g_{\mu\nu}$, which also raises the indices of
$H_{\mu\nu}$. This choice preserves background diffeomorphism
invariance of the radiation effective
action~\cite{GoldbergerRothstein2006,Leibovich2020}.
Because $S_{\rm GF}$ depends on the radiation background,
its variation with respect to $\bar h_{\mu\nu}$ generates
a radiation--potential--potential vertex that must be
included together with the Einstein--Hilbert vertices.

With this gauge choice, the effective action follows from
the potential-mode functional integral,
\begin{equation}
\Gamma[\bar h,\bar\varphi;x_A]
=-\ii\log\!\left[
\int\mathcal DH\,\mathcal D\Phi\,
e^{\ii(S+S_{\rm GF})}
\right].
\label{eq:effective-action}
\end{equation}
This functional integral removes the potential modes
while retaining the radiation fields as dynamical
degrees of freedom. Their subsequent integration
with a causal prescription determines radiation reaction.

Expanding this functional to first order in the radiation
fields defines the composite gravitational and scalar
sources, $T^{\mu\nu}$ and $J$:
\begin{equation}
\Gamma\supset
-\frac{\kappa}{2}\int\dd^{d+1}x\,
T^{\mu\nu}\bar h_{\mu\nu}
-\kappa\int\dd^{d+1}x\,J\bar\varphi.
\label{eq:source-action}
\end{equation}
The gravitational source includes contributions from both
the bodies and their potential fields, and therefore
represents the stress-energy pseudotensor of the composite
system. In the conservative matching problem, with no
external scalar gradient, background diffeomorphism
invariance implies $\partial_\mu T^{\mu\nu}=0$ on the
conservative equations of motion. Once radiation reaction
is included, the exchange of energy and momentum with
the radiation fields must also be accounted for.
The corresponding balance relations are discussed
in Sec.~\ref{sec:dynamics}.

At wavelengths much larger than the source, the spatial
structure of these couplings is resolved through a multipole
expansion. The resulting multipole couplings are collected
in the action~\cite{Thorne1980,GoldbergerRoss2010,Ross2012,Amalberti2024d}

\begin{widetext}
\begin{align}
S_{\rm rad}={}&
-\int\dd\tau\left[
M_{\rm bin}
+\frac12 L^{ab}\omega_{ab\mu}u^\mu
+\sum_{\ell\ge2}\left(
\frac{I^L}{\ell!}\nabla_{L-2}
E_{i_{\ell-1}i_\ell}
-\frac{2\ell J^L}{(\ell+1)!}\nabla_{L-2}
B_{i_{\ell-1}i_\ell}
\right)\right]
-\kappa\int\dd t
\sum_{\ell\ge0}\frac{Q^L}{\ell!}
\partial_L\bar\varphi.
\label{eq:rad-action}
\end{align}
\end{widetext}
The first two terms describe the coupling of the binary's
total mass-energy $M_{\rm bin}$, including its binding
energy, and angular momentum $L^{ab}$ to the radiation
background. Here $\tau$ and $u^\mu$ are the proper time and
four-velocity of the effective binary worldline,
$\omega_{ab\mu}$ is the background spin connection, and
$a,b$ label a local orthonormal frame.

The remaining terms resolve the source into its mass,
current, and scalar multipole moments, denoted by $I^L$,
$J^L$, and $Q^L$, respectively, where $L=i_1\cdots i_\ell$
is a spatial multi-index of rank $\ell$.
The mass and current moments couple
to the electric and magnetic parts of the background Weyl
tensor, $E_{ij}$ and $B_{ij}$, with $\nabla_{L-2}$ denoting
the corresponding product of $\ell-2$ spatial derivatives.
The coefficient $2\ell/(\ell+1)!$ specifies the normalization
of the current moments used in the energy flux below.
The scalar moments couple to spatial derivatives of
$\bar\varphi$, with
$\partial_L=\partial_{i_1}\cdots\partial_{i_\ell}$. All these fields are
evaluated at the spatial origin in the binary rest frame.

Matching the source couplings in Eq.~\eqref{eq:source-action}
to the multipole action in Eq.~\eqref{eq:rad-action}
determines the source moments. These specify the couplings
of the binary to radiation, whereas the radiative moments
that characterize the asymptotic waveform also include
nonlinear propagation effects~\cite{Almeida:2021jyt,Almeida:2024lbv}. In Sec.~\ref{sec:flux}, we
combine our source moments with the nonlinear radiative-moment
relation adopted from Ref.~\cite{BBT2022}, retaining the
contributions required for the instantaneous flux.
An independent EFT derivation of that relation is left
for future work.

\subsection{Fourier expansion and post-Newtonian orders}

The matching described above becomes explicit by expanding
the composite sources in the radiation momentum. Each Taylor
coefficient corresponds to a spatial moment of the source,
as follows from the partial Fourier transform
\begin{widetext}
\begin{equation}
T^{\mu\nu}(t,\vk)
=\int\dd^d\vx\,e^{-\ii\vk\cdot\vx}T^{\mu\nu}(t,\vx)
=\sum_{n\ge0}\frac{(-\ii)^n}{n!}
k_{i_1}\cdots k_{i_n}
\int\dd^d\vx\,x^{i_1}\cdots x^{i_n}
T^{\mu\nu}(t,\vx).
\label{eq:fourier-source}
\end{equation}
\end{widetext}
We use the same convention for the scalar source $J(t,\vk)$.
In the diagrammatic calculation, the endpoint factors
$e^{-\ii\vk\cdot\vx_A}$ are kept explicit. The Taylor
expansion must nevertheless include the full momentum
dependence of each contribution: every explicit factor
of $k_i$ counts toward the total degree, whether it comes
from the endpoint phase or from the remaining amplitude.

The required Taylor degree is tied to the PN order of the
source component. At leading order,
$T^{00}\sim m$, $T^{0i}\sim mv$, and $T^{ij}\sim mv^2$.
The spatial trace, defined by
$T^{ll}\equiv\delta_{ij}T^{ij}$, also starts at order $mv^2$,
whereas $J$ starts at velocity order $v^0$, with coefficients
proportional to the scalar charges. To distinguish source
orders from moment orders, we use a subscript $(v^n)$ for
the absolute velocity order of a source and square brackets
$[n]$ for the relative $n$PN coefficient of a moment.
Table~\ref{tab:orders} summarizes the source orders and
momentum expansions required for the integer-PN matching.
The leading radiation-reaction and nonlinear propagation
contributions are treated separately in
Secs.~\ref{sec:dynamics} and \ref{sec:flux}.

\begin{table}[!t]
\caption{Absolute source orders and maximal total powers
of radiation momentum required for the integer-PN source moments.
The velocity orders and momentum degrees in each row
are paired in the order shown.}
\label{tab:orders}
\begin{ruledtabular}
\begin{tabular}{lll}
Source & Powers of $v$ & Powers of $|\vk|$\\
$T^{00}$ & $0,2,4$ & $6,4,2$\\
$T^{0i}$ & $1,3$ & $5,3$\\
$T^{ll}$ & $2,4$ & $4,2$\\
$T^{ij}$ & $2$ & $4$\\
$J$ & $0,2,4$ & $5,3,1$
\end{tabular}
\end{ruledtabular}
\end{table}

The origin of these different truncations is illustrated
by the mass quadrupole. Its source representation combines
spatial moments with time derivatives, so a fixed PN order
receives contributions from several source components and
Taylor degrees. In three spatial dimensions, the all-orders
expression is~\cite{Ross2012,Leibovich2020}
\begin{widetext}
\begin{align}
I^{ij}={}&
\sum_{p=0}^\infty
\frac{5!!}{(2p)!!(5+2p)!!}
\STF_{ij}\int\dd^3\vx\,x^ix^j(x^2)^p
\left[
\left(1+\frac{2p(3+p)}3\right)
\partial_t^{2p}T^{00}
+\left(1+\frac p3\right)
\partial_t^{2p}T^{ll}
\right.
\nonumber\\
&\left.\hspace{27mm}
-\frac43\left(1+\frac p2\right)
x_a\partial_t^{2p+1}T^{0a}
+\frac16x_ax_b\partial_t^{2p+2}T^{ab}
\right].
\label{eq:I2-all-orders}
\end{align}
\end{widetext}
The symbol STF${}_{ij\dots k}$ represents the symmetric trace-free (STF) projection of the tensor that follows it with respect to the indices $i,j,\dots,k$.
To display its PN expansion, we introduce the shorthand
$\langle F\rangle^{ij}
=\STF_{ij}\int\dd^3\vx\,x^ix^jF$.
Collecting the integer-PN terms through 2PN gives
\begin{widetext}
\begin{align}
I^{ij}_{[0]}
={}&\langle T^{00}_{(v^0)}\rangle^{ij},
\label{eq:I2-order0}\\
I^{ij}_{[1]}
={}&\left\langle
T^{00}_{(v^2)}+T^{ll}_{(v^2)}
-\frac43\partial_t(x_aT^{0a}_{(v)})
+\frac{11}{42}\partial_t^2
(x^2T^{00}_{(v^0)})
\right\rangle^{ij},
\label{eq:I2-order1}\\
I^{ij}_{[2]}
={}&\left\langle
T^{00}_{(v^4)}+T^{ll}_{(v^4)}
-\frac43\partial_t(x_aT^{0a}_{(v^3)})
+\frac{11}{42}\partial_t^2(x^2T^{00}_{(v^2)})
+\frac{2}{21}\partial_t^2(x^2T^{ll}_{(v^2)})
+\frac16\partial_t^2(x_ax_bT^{ab}_{(v^2)})
\right.
\nonumber\\
&\left.\hspace{10mm}
-\frac17\partial_t^3(x^2x_aT^{0a}_{(v)})
+\frac{23}{1512}\partial_t^4(x^4T^{00}_{(v^0)})
\right\rangle^{ij}
+I^{ij}_{[1]}[\va_{[1]}].
\label{eq:I2-order2}
\end{align}
\end{widetext}

The time derivatives in these expressions generate
accelerations, which must also be reduced consistently
in the PN expansion. The final term in
Eq.~\eqref{eq:I2-order2},
$I^{ij}_{[1]}[\va_{[1]}]$, denotes the 2PN correction
induced by inserting the 1PN acceleration into the
time derivatives of the 1PN moment. All other acceleration
reductions in that equation use the Newtonian equations
of motion, so this contribution is included only once.

These reductions determine the integer-PN source moments.
Inserting the 1.5PN reaction force into the 1PN terms of
the source quadrupole would first change that moment at
2.5PN. At the flux orders retained here, the reaction force
instead contributes through derivatives of the leading
moments, as we discuss in Sec.~\ref{sec:results}.

These expressions also explain the tensor-source entries
in Table~\ref{tab:orders}. The largest spatial degree is
six, arising from $x^4x^ix^jT^{00}_{(v^0)}$.
The corresponding contribution therefore requires the
source through sixth order in $\vk$.
By contrast, the uncontracted spatial tensor $T^{ij}$
is needed only at order $v^2$, since its contribution
already carries two time derivatives. Its trace enters
with fewer derivatives and is required through $v^4$.

The scalar moments follow the same logic. In generic spatial dimension $d$, their source representation is
\begin{equation}
Q_L=
\sum_{p\ge0}
\frac{\Gamma(\ell+d/2)}
{2^{2p}p!\Gamma(\ell+p+d/2)}
\partial_t^{2p}
\int\dd^d\vx\,(x^2)^p x_{\langle L\rangle}J.
\label{eq:scalar-series}
\end{equation}
Here $\Gamma(z)$ is the Euler gamma function, and
$x_{\langle L\rangle}$ denotes the symmetric trace-free
product of $\ell$ spatial coordinates. In this expression, 
each term contains
a spatial polynomial of degree $\ell+2p$, and therefore
requires the source through that degree in $\vk$.

For the scalar dipole through relative 2PN order, this
relation selects the linear Taylor coefficient of
$J_{(v^4)}$, the cubic coefficient of $J_{(v^2)}$, and
the quintic coefficient of $J_{(v^0)}$, giving the final
row of Table~\ref{tab:orders}. At the scalar flux accuracy
retained here, the quadrupole is needed only through 1PN;
its 2PN correction contributes beyond that order.
The dissipative center-of-mass transformation adds a
1.5PN term to the scalar dipole, obtained from its leading
source rather than an additional entry in Table~\ref{tab:orders}.
Extending the integer-PN scalar flux further requires
additional source terms, including $J_{(v^6)}$.
The nonlinear propagation terms included in the present
instantaneous flux are specified in Sec.~\ref{sec:flux};
tail integrals remain outside that instantaneous sector.

\section{Radiation sources}\label{sec:calculation}

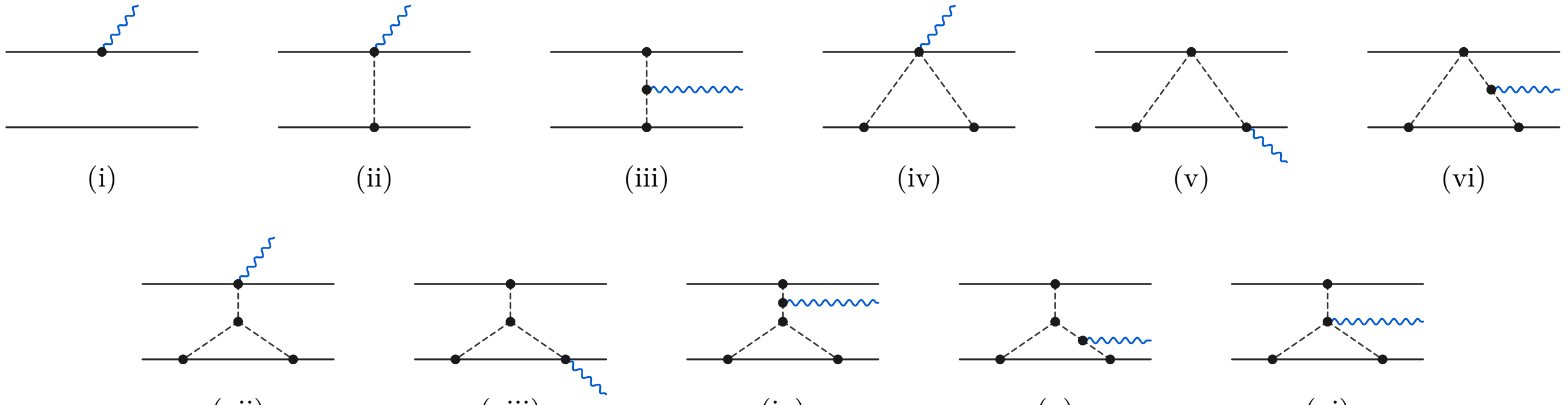
\begin{figure*}[t]
\centering
\scalebox{1}{%
\begin{tikzpicture}[
  x=0.65cm,y=0.65cm,
  world/.style={
    draw=black!85,line width=0.60pt,line cap=round
  },
  pot/.style={
    draw=black!80,line width=0.52pt,
    dash pattern=on 1.7pt off 1.3pt,line cap=round
  },
  rad/.style={
    draw=linkblue,line width=0.60pt,decorate,
    decoration={
      snake,amplitude=0.9pt,segment length=3.8pt,
      pre length=1pt,post length=0pt
    },
    line cap=round
  },
  vertex/.style={
    circle,fill=black!90,draw=none,inner sep=1.1pt
  },
  panel/.style={
    font=\small,inner sep=0pt,anchor=north
  }
]

\foreach \xx/\yy/\lab in {
  0/0/i,
  4.7/0/ii,
  9.4/0/iii,
  14.1/0/iv,
  18.8/0/v,
  23.5/0/vi,
  2.35/-4/vii,
  7.05/-4/viii,
  11.75/-4/ix,
  16.45/-4/x,
  21.15/-4/xi\!\!
}{
  \begin{scope}[shift={(\xx,\yy)}]
    \draw[world] (-1.65,0.65)--(1.65,0.65);
    \draw[world] (-1.65,-0.65)--(1.65,-0.65);
    \node[panel] at (0,-1.3) {(\lab)};
  \end{scope}
}

\begin{scope}[shift={(0,0)}]
  \draw[rad] (0,0.65)--(0.62,1.45);
  \node[vertex] at (0,0.65) {};
\end{scope}

\begin{scope}[shift={(4.7,0)}]
  \draw[pot] (0,0.65)--(0,-0.65);
  \draw[rad] (0,0.65)--(0.62,1.45);
  \node[vertex] at (0,0.65) {};
  \node[vertex] at (0,-0.65) {};
\end{scope}

\begin{scope}[shift={(9.4,0)}]
  \draw[pot] (0,0.65)--(0,0);
  \draw[pot] (0,0)--(0,-0.65);
  \draw[rad] (0,0)--(1.65,0);
  \node[vertex] at (0,0.65) {};
  \node[vertex] at (0,0) {};
  \node[vertex] at (0,-0.65) {};
\end{scope}

\begin{scope}[shift={(14.1,0)}]
  \draw[pot] (0,0.65)--(-0.95,-0.65);
  \draw[pot] (0,0.65)--(0.95,-0.65);
  \draw[rad] (0,0.65)--(0.62,1.45);
  \node[vertex] at (0,0.65) {};
  \node[vertex] at (-0.95,-0.65) {};
  \node[vertex] at (0.95,-0.65) {};
\end{scope}

\begin{scope}[shift={(18.8,0)}]
  \draw[pot] (0,0.65)--(-0.95,-0.65);
  \draw[pot] (0,0.65)--(0.95,-0.65);
  \draw[rad] (0.95,-0.65)--(1.65,-1.25);
  \node[vertex] at (0,0.65) {};
  \node[vertex] at (-0.95,-0.65) {};
  \node[vertex] at (0.95,-0.65) {};
\end{scope}

\begin{scope}[shift={(23.5,0)}]
  \draw[pot] (0,0.65)--(-0.95,-0.65);
  \draw[pot] (0,0.65)--(0.475,0);
  \draw[pot] (0.475,0)--(0.95,-0.65);
  \draw[rad] (0.475,0)--(1.65,0);
  \node[vertex] at (0,0.65) {};
  \node[vertex] at (-0.95,-0.65) {};
  \node[vertex] at (0.475,0) {};
  \node[vertex] at (0.95,-0.65) {};
\end{scope}

\begin{scope}[shift={(2.35,-4)}]
  \draw[pot] (0,0.65)--(0,0);
  \draw[pot] (0,0)--(-0.95,-0.65);
  \draw[pot] (0,0)--(0.95,-0.65);
  \draw[rad] (0,0.65)--(0.62,1.45);
  \node[vertex] at (0,0.65) {};
  \node[vertex] at (0,0) {};
  \node[vertex] at (-0.95,-0.65) {};
  \node[vertex] at (0.95,-0.65) {};
\end{scope}

\begin{scope}[shift={(7.05,-4)}]
  \draw[pot] (0,0.65)--(0,0);
  \draw[pot] (0,0)--(-0.95,-0.65);
  \draw[pot] (0,0)--(0.95,-0.65);
  \draw[rad] (0.95,-0.65)--(1.65,-1.25);
  \node[vertex] at (0,0.65) {};
  \node[vertex] at (0,0) {};
  \node[vertex] at (-0.95,-0.65) {};
  \node[vertex] at (0.95,-0.65) {};
\end{scope}

\begin{scope}[shift={(11.75,-4)}]
  \draw[pot] (0,0.65)--(0,0.325);
  \draw[pot] (0,0.325)--(0,0);
  \draw[pot] (0,0)--(-0.95,-0.65);
  \draw[pot] (0,0)--(0.95,-0.65);
  \draw[rad] (0,0.325)--(1.65,0.325);
  \node[vertex] at (0,0.65) {};
  \node[vertex] at (0,0.325) {};
  \node[vertex] at (0,0) {};
  \node[vertex] at (-0.95,-0.65) {};
  \node[vertex] at (0.95,-0.65) {};
\end{scope}

\begin{scope}[shift={(16.45,-4)}]
  \draw[pot] (0,0.65)--(0,0);
  \draw[pot] (0,0)--(-0.95,-0.65);
  \draw[pot] (0,0)--(0.475,-0.325);
  \draw[pot] (0.475,-0.325)--(0.95,-0.65);
  \draw[rad] (0.475,-0.325)--(1.65,-0.325);
  \node[vertex] at (0,0.65) {};
  \node[vertex] at (0,0) {};
  \node[vertex] at (-0.95,-0.65) {};
  \node[vertex] at (0.475,-0.325) {};
  \node[vertex] at (0.95,-0.65) {};
\end{scope}

\begin{scope}[shift={(21.15,-4)}]
  \draw[pot] (0,0.65)--(0,0);
  \draw[pot] (0,0)--(-0.95,-0.65);
  \draw[pot] (0,0)--(0.95,-0.65);
  \draw[rad] (0,0)--(1.65,0);
  \node[vertex] at (0,0.65) {};
  \node[vertex] at (0,0) {};
  \node[vertex] at (-0.95,-0.65) {};
  \node[vertex] at (0.95,-0.65) {};
\end{scope}

\end{tikzpicture}
}
\caption{Topology classes contributing to the linear radiation
sources at the orders retained in this work. The upper and
lower solid horizontal lines represent the worldlines of
bodies $A$ and $B$, respectively, while straight dashed
internal lines denote potential fields and the wavy external
leg denotes radiation. Panel (i) describes direct emission
without potential exchange; panels (ii)--(iii) contribute
at order $G$, and panels (iv)--(xi) at order $G^2$.
The classes distinguish both the radiation attachment and
the distribution of vertices between the two worldlines.
Each panel represents the allowed tensor and scalar field
assignments, together with their body-exchanged counterparts
and the velocity and retardation insertions required by
the PN expansion. Bulk interactions arise from the vertices
$H^3$, $H\Phi^2$, $H^4$, $H^2\Phi^2$,
with one field replaced by its radiation counterpart
where appropriate.}
\label{fig:topologies}
\end{figure*}

We now construct the composite sources introduced in
Eq.~\eqref{eq:source-action}. Their contributions are obtained
from connected diagrams with one external radiation field
and potential fields consistent with the PN counting.
Figure~\ref{fig:topologies} collects the eleven
topology classes required for this potential-mode matching.
Radiation reaction and nonlinear propagation involve the
radiation fields and are treated in
Secs.~\ref{sec:dynamics} and \ref{sec:flux}, respectively.

The position of the radiation attachment and the distribution
of vertices between the two worldlines distinguish these
classes; only a simultaneous exchange $A\leftrightarrow B$
is implicit. Within each class, assigning tensor or scalar
fields to the lines and expanding the vertices in velocities
generates the corresponding gravitational and scalar source
terms. The scalar-dependent worldline masses supply mixed
contact vertices, while interactions from the bulk scalar
action involve derivatives of the scalar and at least one
gravitational field.

To extract the momentum coefficients needed for the
multipoles, we expand the potential propagators in the small
radiation momentum. For a shifted potential denominator,
this gives
\begin{equation}
\frac1{(\vq+\vk)^2}
=\frac1{\vq^2}
-\frac{2\vq\cdot\vk+\vk^2}{(\vq^2)^2}
+\frac{(2\vq\cdot\vk+\vk^2)^2}{(\vq^2)^3}
+\cdots .
\label{eq:potential-spatial-expansion}
\end{equation}
Since $|\vk|/|\vq|\sim v/c$, the ratio
$\vq\cdot\vk/\vq^2$ carries one power of $v/c$,
whereas $\vk^2/\vq^2$ carries two. This expansion
therefore implements the same long-wavelength counting
as the Taylor expansion of the external radiation field.
Because a single numerator in
Eq.~\eqref{eq:potential-spatial-expansion} contains several
momentum degrees, the truncation must be imposed on the
total power of $\vk$. At quadratic order, for example,
the last displayed term contributes
$4(\vq\cdot\vk)^2/(\vq^2)^3$, while its cubic and quartic
pieces lie beyond the truncation.

The frequency dependence of the potential propagators
is treated in the same way:
\begin{equation}
\frac1{\vq^2-(q^0)^2}
=\frac1{\vq^2}
+\frac{(q^0)^2}{(\vq^2)^2}
+\frac{(q^0)^4}{(\vq^2)^3}
+\cdots .
\label{eq:potential-time-expansion}
\end{equation}
This temporal expansion organizes retardation corrections
in powers of $v^2$. Both expansions are performed before
the potential integrations, following the procedure of
Ref.~\cite{Leibovich2020}. The accompanying momentum
numerators are determined by the action, including the
gauge-fixing term, and frequency insertions are expressed
as time derivatives acting on the vertices and endpoint
phases.

After these expansions, the potential integrals are evaluated
in dimensional regularization. Tensor integrals are reduced
to scalar Fourier integrals and their
derivatives~\cite{GilmoreRoss2008,KolSmolkin2008,FoffaSturani2014,PortoRothstein2017,Foffa2019ren}.
In carrying out the subsequent multipole projection, terms
proportional to $\vk^2$ are retained until all trace
derivatives in the multipole operators have acted.
The complete spatial and time-derivative operators,
applied to the summed sources, give vanishing pole
residues for the moments retained here. This cancellation
occurs at the level of the projected moments; individual
bare source components need not be finite.
Accordingly, the projection is performed in analytically
continued spatial dimension before taking $d\to3$.

\begin{table}[!t]
\caption{Number of potential-mode matching contributions at each absolute
velocity order, within the radiation-momentum truncations of
Table~\ref{tab:orders}. Distinct velocity and retardation
insertions are counted separately, as are distinct diagrams
related by $A\leftrightarrow B$. Entries in each row correspond
in order. The trace $T^{ll}$ is listed separately from the open $T^{ij}$.
Thus, their entries are not independent
sets of graphs.}
\label{tab:diagram-counts}
\begin{ruledtabular}
\begin{tabular}{lcc}
Source & Powers of $v$ & Contributions \\
\hline
$T^{00}$ & $0,2,4$ & $2,8,77$ \\
$T^{0i}$ & $1,3$   & $2,12$ \\
$T^{ij}$ & $2$     & $4$ \\
$T^{ll}$ & $2,4$   & $4,47$ \\
$J$      & $0,2,4$ & $2,6,54$
\end{tabular}
\end{ruledtabular}
\end{table}

We perform the tensor algebra using the \texttt{xAct}
packages~\cite{xAct,MartinGarcia2008xPerm}.
Table~\ref{tab:diagram-counts} summarizes the diagrammatic
contributions to each source component at the velocity
and radiation-momentum orders required for the retained
multipoles.
These counts refer to the source calculation; the radiative
integration generating the dipolar reaction force and
the adopted nonlinear propagation terms are not included
in Table~\ref{tab:diagram-counts}.

The simplest source terms arise from direct emission by
the point particles. At leading order, they are
\begin{align}
T^{00}_{(v^0)}(t,\vk)
&=\sum_A m_A e^{-\ii\vk\cdot\vx_A},
\nonumber\\
T^{0i}_{(v)}(t,\vk)
&=\sum_A m_Av_A^i e^{-\ii\vk\cdot\vx_A},
\nonumber\\
T^{ij}_{{\rm pp},(v^2)}(t,\vk)
&=\sum_A m_Av_A^iv_A^j e^{-\ii\vk\cdot\vx_A},
\nonumber\\
J_{(v^0)}(t,\vk)
&=\sum_A m_A\alpha_A e^{-\ii\vk\cdot\vx_A}.
\label{eq:leading-sources}
\end{align}
The subscript ${\rm pp}$ identifies the point-particle
contribution to the spatial tensor, since potential fields
also contribute to the full $T^{ij}_{(v^2)}$.

At the next order, the scalar source illustrates how
velocity corrections to direct emission combine with
potential exchange:
\begin{widetext}
\begin{align}
J_{(v^2)}(t,\vk)
={}&-\frac12\sum_A
m_A\alpha_A v_A^2e^{-\ii\vk\cdot\vx_A}
\nonumber\\
&-\frac{Gm_Am_B}{r}
\left[
\left(\alpha_A+
\left(\alpha_A^2+\beta_A\right)\alpha_B\right)
e^{-\ii\vk\cdot\vx_A}
+(A\leftrightarrow B)
\right].
\label{eq:J2}
\end{align}
\end{widetext}
The first term is the velocity correction to the
worldline source, while the second contains the exchange
contributions and their dependence on the scalar response.
At higher orders, all the topologies in
Fig.~\ref{fig:topologies} enter the source construction.
We organize their combined contributions into the
multipole moments presented  in Sec.~\ref{sec:results}.
\section{Orbital dynamics and balance laws}\label{sec:dynamics}

The time derivatives of the source multipoles depend on
the orbital motion. At the flux orders considered here,
they require both the conservative dynamics through 2PN
and the leading dipolar radiation reaction at 1.5PN.
We present these ingredients in the coordinates used for
the source calculation, then determine the conserved
quantities of the conservative sector and the dissipative
correction to the center-of-mass frame.

\subsection{Conservative acceleration}

We explicitly compute the conservative effective action
following the EFT procedure of Ref.~\cite{Almeida2024},
while imposing the linear harmonic (de Donder) gauge
used for the radiation sources. The details of this
calculation and the resulting action are presented
in Appendix~\ref{app:action}. To relate our result to the nonlinear
harmonic action of Ref.~\cite{Almeida2024}, we write
the Lagrangian as
$L=-mc^2+L_0+c^{-2}L_1+c^{-4}L_{2,\rm lin}$.
The two descriptions agree at Newtonian and 1PN orders,
while their 2PN coefficients satisfy
\begin{align}
L_{2,\rm lin}
&=L_{2,\rm harm}+\delta L_2,
\label{eq:action-relation}\\
\delta L_2
&=\frac{2G^2m_Am_Bm}{r^2}
\left(v^2-2\dot r^2-\frac{G_{AB}m}{r}\right).
\label{eq:delta-L}
\end{align}
The explicit factor $G^2$ reflects the metric gauge
transformation, whereas $G_{AB}$ enters through the
Newtonian interaction. This distinction must be preserved:
replacing every occurrence of $G$ by $G_{AB}$ would
change the action.

The corresponding change of coordinates is generated by
the 2PN displacement
\begin{align}
\vx_A^{\rm lin}
&=\vx_A^{\rm harm}+c^{-4}\boldsymbol\xi_A,
\nonumber\\
\boldsymbol\xi_A
&=-2G^2m_Bm\frac{\vr}{r^2},
\qquad
\boldsymbol\xi_B
=2G^2m_Am\frac{\vr}{r^2}.
\label{eq:gauge-shift}
\end{align}
This displacement allows us to compare the accelerations
obtained in the two gauges. 

To derive the accelerations, 
we first vary the generalized Lagrangian, retaining its
acceleration dependence in the Euler--Lagrange equations,
\begin{equation}
\frac{\partial L}{\partial x_A^i}
-\frac{\dd}{\dd t}\frac{\partial L}{\partial v_A^i}
+\frac{\dd^2}{\dd t^2}\frac{\partial L}{\partial a_A^i}
=0,
\label{eq:EL}
\end{equation}
and only then reduce the result using the lower-order
equations of motion. At 2PN, this reduction includes
the 1PN acceleration induced by the lower-order terms.

The conservative acceleration in these coordinates takes
the compact form

\begin{align}
\va_{A,\mathrm{cons}}
={}&-\frac{G_{AB}m_B}{r^3}\vr
\nonumber\\
&+\sum_{n=1}^2
\frac{\mathscr A_{A,n}\vr+\mathscr B_{A,n}\vv}{c^{2n}}
+\order(c^{-6}).
\label{eq:acceleration}
\end{align}
The coefficient functions $\mathscr A_{A,n}$ and
$\mathscr B_{A,n}$ depend on $r$, $\vr\cdot\vv_A$,
$\vr\cdot\vv_B$, $v_A^2$, $v_B^2$, and $\vv_A\cdot\vv_B$,
together with the masses and scalar responses.
Their explicit expressions are given in
Appendix~\ref{app:acceleration}.

Applying Eq.~\eqref{eq:gauge-shift} to this result
reproduces the harmonic-coordinate acceleration of
Ref.~\cite{Almeida2024}. At 2PN, the transformation involves
both the second time derivative of the displacement and
the change in the argument of the Newtonian force:
\begin{align}
\va_{A,[2]}^{\rm lin}
=\va_{A,[2]}^{\rm harm}
+D_0^2\boldsymbol\xi_A
-(\boldsymbol\xi_A-\boldsymbol\xi_B)
\cdot\boldsymbol\nabla_r\va_{A,[0]}.
\label{eq:acc-shift}
\end{align}
Here $D_0$ denotes time evolution with the Newtonian
equations of motion, and $\boldsymbol\nabla_r$
differentiates with respect to the relative position.
Together, these two terms account for the coordinate
dependence of the 2PN acceleration.

\subsection{Leading dipolar radiation reaction}\label{sec:dipole-reaction}

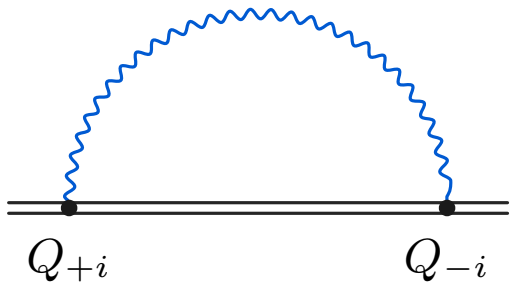
\begin{figure}[t]
\centering
\begin{tikzpicture}[
  x=1cm,y=1cm,
  world/.style={
    draw=black!85,line width=0.60pt,line cap=round
  },
  rad/.style={
    draw=linkblue,line width=0.60pt,decorate,
    decoration={snake,amplitude=0.9pt,segment length=3.8pt,
      pre length=1pt,post length=1pt},
    line cap=round
  },
  vertex/.style={
    circle,fill=black!90,draw=none,inner sep=1.1pt
  },
  source/.style={font=\small,inner sep=0pt,anchor=north}
]
  \draw[world] (-1.65,0.035)--(1.65,0.035);
  \draw[world] (-1.65,-0.035)--(1.65,-0.035);

  \draw[rad] (-1.25,0) arc[start angle=180,end angle=0,radius=1.25];
  \node[vertex] at (-1.25,0) {};
  \node[vertex] at (1.25,0) {};

  \node[source] at (-1.25,-0.20) {$Q_{+i}$};
  \node[source] at (1.25,-0.20) {$Q_{-i}$};
\end{tikzpicture}
\caption{Scalar dipole contribution to the effective action
$S^{\rm dip}$. The double horizontal line represents
the binary as a composite radiation source. The black vertices
mark the $Q_{+i}$ and $Q_{-i}$ dipole insertions in the doubled-history
formalism, connected by a scalar radiation propagator (blue wavy
line).}
\label{fig:dipole-reaction}
\end{figure}

The scalar dipole provides the leading dissipative force.
With the canonical radiation field defined in
Sec.~\ref{sec:setup}, its coupling is
$-\kappa Q_i\partial_i\bar\vartheta/\sqrt{8}$.
To integrate out this field with outgoing boundary
conditions, we use the causal doubled-history
formalism~\cite{GalleyTiglio2009,GalleyLeibovich2012}.
For two worldline histories, we define
$\vx_{A+}=(\vx_{A1}+\vx_{A2})/2$ and
$\vx_{A-}=\vx_{A1}-\vx_{A2}$, with analogous definitions
for $Q_{+i}$ and $Q_{-i}$. 
At the required order,
$Q_i=\sum_A m_A\alpha_A x_A^i$.
The contraction of two dipole insertions through the
causal scalar propagator is shown in
Fig.~\ref{fig:dipole-reaction}. The double horizontal
line represents the binary as a composite radiation
source.

Integrating out the scalar radiation field gives
the dipolar radiation-reaction effective action,
\begin{equation}
S^{\rm dip}
=\frac{G}{3c^3}\int\dd t\,Q_{-i}Q_{+i}^{(3)}.
\label{eq:dipole-influence}
\end{equation}
A parenthesized superscript denotes a time derivative.
The coefficient follows from the retarded dipole kernel:
at $c=1$, with Fourier convention $e^{-\ii\omega t}$,
dimensional regularization yields
\begin{align}
\mathcal K_{ij}(\omega)
&=\frac{\kappa^2}{8}
\int\frac{\dd^3\vq}{(2\pi)^3}
\frac{q_iq_j}{\vq^2-(\omega+\ii0)^2}
\nonumber\\
&=\frac{\ii G\omega^3}{3}\delta_{ij}.
\label{eq:dipole-retarded-kernel}
\end{align}
Here $\omega$ is the frequency, $\vq$ is an integration
momentum, and $\mathcal K_{ij}$ denotes the scalar kernel.
The infinitesimal prescription selects the retarded
solution.

Varying with respect to $\vx_{A-}$ before taking the
physical limit $\vx_{A-}=0$ gives
\begin{equation}
a_{A,\mathrm{RR}}^i
=\frac{G\alpha_A}{3c^3}Q_i^{(3)}.
\label{eq:reaction-unreduced}
\end{equation}
The label $\mathrm{RR}$ denotes radiation reaction.
Reducing the derivative with the Newtonian equations
then gives the individual accelerations
\begin{align}
\va_{A,\mathrm{RR}}
&=\frac{G_{AB} G m_Am_B}{3c^3r^3}
\alpha_A(\alpha_A-\alpha_B)
(3\dot r\vn-\vv),\nonumber\\
\va_{B,\mathrm{RR}}
&=\frac{G_{AB} G m_Am_B}{3c^3r^3}
\alpha_B(\alpha_A-\alpha_B)
(3\dot r\vn-\vv).
\label{eq:individual-reaction}
\end{align}
Both expressions use $\vr=\vx_A-\vx_B$ and
$\vv=\vv_A-\vv_B$. Their difference is
\begin{equation}
\va_{\mathrm{RR}}
=\frac{G_{AB}G m\mu(\alpha_A-\alpha_B)^2}
{3c^3r^3}(3\dot r\vn-\vv).
\label{eq:relative-reaction}
\end{equation}
These results reproduce the leading harmonic-coordinate
reaction force of Refs.~\cite{MirshekariWill2013,BBT2022}
after translating the scalar parameters. 
The acceleration used below is therefore
\begin{equation}
\va_A=\va_{A,\mathrm{cons}}+\va_{A,\mathrm{RR}}
+\order(c^{-5}),
\label{eq:acceleration-with-reaction}
\end{equation}
where the conservative part includes the 2PN terms of
Eq.~\eqref{eq:acceleration}, while the omitted dissipative
terms begin at 2.5PN order.

\subsection{Conservative charges and center of mass}

The same conservative action also determines the conserved
quantities through its spacetime symmetries, independently
of the multipole matching. To apply Noether's theorem
in the presence of acceleration dependence, we introduce
the generalized momenta
\begin{equation}
\mathbf p_A
=\frac{\partial L}{\partial\vv_A}
-\frac{\dd}{\dd t}\frac{\partial L}{\partial\va_A},
\qquad
\mathbf q_A
=\frac{\partial L}{\partial\va_A}.
\label{eq:generalized-momenta}
\end{equation}
The auxiliary momentum $\mathbf q_A$ accounts for the
acceleration dependence of the Lagrangian and is unrelated
to the potential momentum $\vq$. In terms of these
quantities, time translations, spatial translations,
and rotations yield the energy, linear momentum,
and angular momentum, respectively:
\begin{align}
E
&=\sum_A
\left(\mathbf p_A\cdot\vv_A+\mathbf q_A\cdot\va_A\right)-L,
\nonumber\\
\mathbf P
&=\sum_A\mathbf p_A,
\qquad
\mathbf L
=\sum_A
\left(\vx_A\times\mathbf p_A+\vv_A\times\mathbf q_A\right).
\label{eq:Noether-charges}
\end{align}

Lorentz boosts complete this construction by determining
the mass dipole $\boldsymbol{\mathcal G}$ and the conserved
boost vector
$\mathbf K=\boldsymbol{\mathcal G}-t\mathbf P$.
Their conservation implies
$\dot{\boldsymbol{\mathcal G}}=\mathbf P$,
which relates the mass dipole to the motion of the
center of mass through
$\mathbf X_{\rm CM}=c^2\boldsymbol{\mathcal G}/E$.
Appendix~\ref{app:Noether} presents the Noether construction
and the explicit charges at the orders tested by the
radiation sources: the energy, linear momentum, and angular
momentum through 1PN, and the mass dipole through 2PN.
These truncations follow from the source orders listed
in Table~\ref{tab:orders}.

Within the conservative sector, the mass dipole fixes
the center-of-mass transformation. In the rest frame,
imposing $\boldsymbol{\mathcal G}=0$ gives the common
displacement from the Newtonian positions:

\begin{align}
\vx_A
&=\frac{m_B}{m}\vr+\delta\vx,
\qquad
\vx_B=-\frac{m_A}{m}\vr+\delta\vx,
\nonumber\\
\delta\vx
&=\left(\frac{f_1}{c^2}+\frac{f_2}{c^4}\right)\vr
+\frac{b_2}{c^4}r\vv.
\label{eq:CM-shift}
\end{align}
The scalar functions $f_1$, $f_2$, and $b_2$ depend on
$Gm/r$, $v^2$, $\dot r$, the mass ratios, and the scalar
responses; their explicit expressions are collected in
Appendix~\ref{app:CM-shift}.

The corresponding velocity shifts follow by differentiating
Eq.~\eqref{eq:CM-shift} consistently in the PN expansion.
In particular, the 2PN velocity shift includes the term
generated by inserting the 1PN acceleration into the
derivative of the 1PN position shift.

\subsection{Dipolar balance and the dissipative center of mass}
\label{sec:dissipative-CM}

The charges above are conserved under the conservative
equations. The reaction force instead gives, at leading
dissipative order, with $\dot E_{\rm mech} = \sum_A m_A \vv_A \cdot \va_{A,{\rm RR}}$, 
\begin{align}
\dot E_{\rm mech} 
&=\frac{G}{3c^3}\dot Q_iQ_i^{(3)},\nonumber\\
E_{\rm Schott}&=-\frac{G}{3c^3}\dot Q_i\ddot Q_i,\nonumber\\
\frac{\dd}{\dd t}(E_{\rm mech}+E_{\rm Schott})
&=-\frac{G}{3c^3}\ddot Q_i\ddot Q_i.
\label{eq:dipole-energy-balance}
\end{align}
Here $E_{\rm mech}$ is the orbital mechanical energy
and $E_{\rm Schott}$ is the reversible near-zone field
contribution to the balance. The right-hand side is
minus the leading outgoing scalar dipole power. The
balance follows from the causal force, independently
checking its sign and normalization.

The corresponding momentum balance involves the constant
leading scalar monopole $Q_{[0]}=m_A\alpha_A+m_B\alpha_B$ and, using 
$\dot {\mathbf P}_{\rm mech} = \sum_A m_A \va_{A,{\rm RR}}$:
\begin{align}
\dot{\mathbf P}_{\rm mech}
&=\frac{GQ_{[0]}}{3c^3}\mathbf Q^{(3)},\nonumber\\
\mathbf P_{\rm Schott}
&=-\frac{GQ_{[0]}}{3c^3}\ddot{\mathbf Q}.
\label{eq:dipole-momentum-balance}
\end{align}
The vector $\mathbf Q$ has components $Q_i$. Pure dipole
emission carries no net linear momentum at this order,
so $\mathbf P_{\rm mech}+\mathbf P_{\rm Schott}$ is
constant at leading dissipative order. The rest-frame
condition must therefore include this field momentum.
Equivalently, the mass dipole receives
$-GQ_{[0]}\dot{\mathbf Q}/(3c^3)$ in addition to its
conservative terms.

Defining the mass-weighted scalar charge
$\bar\alpha=(m_A\alpha_A+m_B\alpha_B)/m$, we now 
set the corrected mass dipole and total rest momentum
to zero: this adds the common position and velocity shifts
\begin{align}
\delta\vx_{\mathrm{RR}}
&=\frac{G\mu(\alpha_A-\alpha_B)\bar\alpha}{3c^3}\vv,
\label{eq:CM-shift-RR}\\
\delta\vv_{\mathrm{RR}}
&=-\frac{G\mu(\alpha_A-\alpha_B)\bar\alpha}{3c^3}
\frac{G_{AB}m}{r^2}\vn.
\label{eq:CM-velocity-RR}
\end{align}
Thus, writing the displacement in Eq.~\eqref{eq:CM-shift}
as $\delta\vx_{\rm cons}$, the positions become
\begin{align}
\vx_A&=\frac{m_B}{m}\vr+\delta\vx_{\rm cons}
+\delta\vx_{\mathrm{RR}}+\order(c^{-5}),\nonumber\\
\vx_B&=-\frac{m_A}{m}\vr+\delta\vx_{\rm cons}
+\delta\vx_{\mathrm{RR}}+\order(c^{-5}).
\label{eq:CM-shift-complete}
\end{align}
The velocity correction in Eq.~\eqref{eq:CM-velocity-RR}
uses Newtonian equations of motion, sufficient at 1.5PN.
These shifts agree with the corresponding harmonic-frame
expressions of Refs.~\cite{MirshekariWill2013,BBT2022}.

\section{Tensor and scalar multipoles}\label{sec:results}

We now combine the radiation sources of
Sec.~\ref{sec:calculation} with the conservative reduction
of Sec.~\ref{sec:dynamics} to obtain the integer-PN source
moments. We then include the dissipative center-of-mass
correction and specify how radiation reaction enters their
time derivatives. Throughout, we retain the linear harmonic
gauge, first giving the moments in an arbitrary reference
frame and then specializing to the center-of-mass frame.

\subsection{Source multipoles in an arbitrary reference frame}

Before applying the source-to-multipole projections,
such as those in Eqs.~\eqref{eq:I2-all-orders}
and \eqref{eq:scalar-series}, we examine the lowest
Fourier coefficients of the sources, which provide
a direct connection with the conserved quantities.
In units $c=1$, these quantities are
\begin{align}
E&=\int\dd^3x\,T^{00},
&P^i&=\int\dd^3x\,T^{0i},
\nonumber\\
\mathcal G^i&=\int\dd^3x\,x^iT^{00},
&L^i&=\epsilon^{ijk}\int\dd^3x\,x^jT^{0k}.
\label{eq:charges-source}
\end{align}
The energy, linear momentum, and angular momentum
obtained from these source integrals agree with the
Noether expressions in Appendix~\ref{app:Noether}
through 1PN. The mass dipole agrees through 2PN and
satisfies $\dot{\boldsymbol{\mathcal G}}=\mathbf P$
with the independently computed 2PN momentum.
These relations hold in the conservative sector and
check its consistency with the source matching.
The additional field contributions required on a
dissipative orbit are given in
Sec.~\ref{sec:dissipative-CM}.

Applying the higher multipole projections then gives,
at leading order in the PN expansion,
\begin{align}
I_{L,[0]}
&=\sum_A m_Ax_A^{\langle L\rangle},
\nonumber\\
J_{L,[0]}
&=\sum_A m_A\STF_L
\left[x_A^{L-1}(\vx_A\times\vv_A)^{i_\ell}\right],
\nonumber\\
Q_{L,[0]}
&=\sum_A m_A\alpha_Ax_A^{\langle L\rangle}.
\label{eq:leading-moments}
\end{align}
The ranks and PN corrections needed for the instantaneous
power are summarized in Table~\ref{tab:moments}, together
with the number of time derivatives acting on each moment
in the flux. The leading scalar monopole,
$Q_{[0]}=\sum_A m_A\alpha_A$, is constant and therefore
does not radiate. Unlike the conserved energy, however,
the full scalar monopole need not be time independent:
its PN corrections can generate monopole radiation
for generic orbits. The first such contribution to
the energy flux occurs one PN order beyond the leading
scalar dipole flux, at the same order as the leading
quadrupole flux~\cite{Lang2015}.

\begin{table*}[t]
\caption{Multipoles required for the instantaneous power.
The PN orders label the coefficients retained in each
moment, and the derivative columns give the number of
time derivatives entering the flux. The constant leading
scalar monopole is omitted because it does not radiate.
The $3/2$ entry is the dissipative center-of-mass correction
to the scalar dipole $Q_i$; nonlinear wave propagation is treated separately in
Sec.~\ref{sec:flux}.}
\label{tab:moments}
\begin{ruledtabular}
\begin{tabular}{lcc@{\qquad\qquad}lcc}
Tensor moment & Relative PN orders & Derivatives
& Scalar moment & Relative PN orders & Derivatives \\
$I_{ij}$     & $0,1,2$ & 3 & $Q$       & $1,2$   & 1 \\
$I_{ijk}$    & $0,1$   & 4 & $Q_i$     & $0,1,3/2,2$ & 2 \\
$I_{ijkl}$   & $0$     & 5 & $Q_{ij}$  & $0,1$   & 3 \\
$J_{ij}$     & $0,1$   & 3 & $Q_{ijk}$ & $0$     & 4 \\
$J_{ijk}$    & $0$     & 4 &           &         &
\end{tabular}
\end{ruledtabular}
\end{table*}

We begin with the mass quadrupole, whose corrections
are required through 2PN. Expressed in terms of the
individual bodies' positions and velocities, the
result through 2PN takes the form
\begin{widetext}
\begin{align}
I^{ij}
={}&\sum_A m_Ax_A^{\langle i}x_A^{j\rangle}
+\frac1{c^2}\sum_A m_A\STF_{ij}
\left[
\left(
\frac{29}{42}v_A^2
-\frac{17}{21}\vx_A\cdot\va_{A,[0]}
\right)x_A^ix_A^j
\right.
\nonumber\\
&\left.\hspace{18mm}
-\frac47(\vx_A\cdot\vv_A)x_A^{(i}v_A^{j)}
+\frac{11}{21}x_A^2v_A^iv_A^j
+\frac{11}{21}x_A^2x_A^{(i}a_{A,[0]}^{j)}
\right]
\nonumber\\
&-\frac{G_{AB}m_Am_B}{c^2r}
\STF_{ij}\left[x_A^ix_A^j+x_B^ix_B^j\right]
+\frac1{c^4}\STF_{ij}
\left[
C_{xx}x_A^ix_A^j
+C_{xy}x_A^{(i}x_B^{j)}
+C_{xv}x_A^{(i}v_A^{j)}
+C_{xw}x_A^{(i}v_B^{j)}
\right.
\nonumber\\
&\left.\hspace{36mm}
+C_{vv}v_A^iv_A^j
+C_{vw}v_A^{(i}v_B^{j)}
+(A\leftrightarrow B)
\right]
+\order(c^{-5}),
\label{eq:I2-general}
\end{align}
\end{widetext}
where parentheses on indices denote symmetrization, as in
$A^{(ij)}=(A^{ij}+A^{ji})/2$, also recalling that 
angle brackets denote the symmetric trace-free projection.
Here $\va_{A,[0]}=-G_{AB}m_B\vr/r^3$ is the
Newtonian acceleration. The six coefficient functions
$C_{xx},\ldots,C_{vw}$, supplied in the ancillary
Mathematica notebook with their structure described in
Appendix~\ref{app:general-I2}, contain the complete
2PN correction, including the terms induced by
lower-order accelerations. 

The scalar moments follow from the corresponding
projections of $J$. For the monopole and dipole,
the first PN corrections are
\begin{widetext}
\begin{align}
Q_{[1]}
={}&\sum_A m_A\alpha_A
\left[
-\frac16v_A^2
+\frac13\vx_A\cdot\va_{A,[0]}
\right]
-\frac{m_Am_B}{r}
\left[
G_{AB}(\alpha_A+\alpha_B)
+G(\beta_A\alpha_B+\beta_B\alpha_A)
\right],
\label{eq:Q0-1PN}\\
Q^i_{[1]}
={}&\sum_A m_A\alpha_A
\left[
\left(
-\frac3{10}v_A^2
+\frac15\vx_A\cdot\va_{A,[0]}
\right)x_A^i
+\frac25(\vx_A\cdot\vv_A)v_A^i
+\frac1{10}x_A^2a_{A,[0]}^i
\right]
\nonumber\\
&-\frac{m_Am_B}{r}
\left[
(G_{AB}\alpha_A+G\beta_A\alpha_B)x_A^i
+(A\leftrightarrow B)
\right].
\label{eq:Q1-1PN}
\end{align}
\end{widetext}
Their complete 2PN coefficients, together with the 1PN mass octupole,
scalar quadrupole, and current quadrupole, are supplied in the ancillary
Mathematica notebook \texttt{ST\_Radiative\_Coefficients.nb}.
Appendices~\ref{app:general-other} and \ref{app:current} specify their
bases and fix the notation used in the ancillary file.
They also display the remaining tensor source operators not shown explicitly here.

\subsection{Center-of-mass expressions}

To express these results in terms of the relative orbit,
we now transform to the center-of-mass frame using
Eqs.~\eqref{eq:CM-shift} and \eqref{eq:CM-shift-complete},
together with the corresponding velocity shifts.
The common dissipative displacement gives no linear
1.5PN correction to the mass quadrupole, since
$\sum_A m_A\vx_{A,[0]}=0$ in the Newtonian center-of-mass frame.
Its expression through 2PN therefore retains three
independent STF tensor structures:

\begin{equation}
I^{ij}
=\mu\STF_{ij}
\left[
\mathcal A r^ir^j
+\mathcal B r^2v^iv^j
+\mathcal C r\dot r\,r^{(i}v^{j)}
\right],
\label{eq:I2-CM}
\end{equation}
with coefficient functions expanded as
$\mathcal A=1+c^{-2}\mathcal A_1+c^{-4}\mathcal A_2$,
$\mathcal B=c^{-2}\mathcal B_1+c^{-4}\mathcal B_2$,
and 
$\mathcal C=c^{-2}\mathcal C_1+c^{-4}\mathcal C_2$.
Collecting the interaction terms with $G_{AB}$ gives
\begin{widetext}
\begin{align}
\mathcal A_1&=\frac{G_{AB}m}{r}\frac{8\nu-5}{7}+\frac{29}{42}(1-3\nu)v^2,\label{eq:CM-I2-A1}\\
\mathcal B_1&=\frac{11}{21}(1-3\nu),\qquad
\mathcal C_1=-\frac47(1-3\nu).\label{eq:CM-I2-B1}
\end{align}
\begin{align}
\mathcal A_2={}&\left(\frac{G_{AB}m}{r}\right)^2
\bigg\{\frac{337\nu^2-1906\nu+653}{252}
+\frac{11}{12}\bar\gamma_{AB}^2+\frac{65-90\nu}{21}\bar\gamma_{AB}
-\frac{(1+\Delta)\delta_A+(1-\Delta)\delta_B}{6}\nonumber\\
&\hspace{20mm}+\frac{5\nu+3+(9\nu-3)\Delta}{14}\bar\beta_A
+\frac{5\nu+3-(9\nu-3)\Delta}{14}\bar\beta_B\bigg\}\nonumber\\
&-\frac{G_{AB}m}{r}\frac{v^2}{756}
\left[4883\nu^2+5947\nu-2021+(4680\nu-1308)\bar\gamma_{AB}\right]\nonumber\\
&-\frac{G_{AB}m}{r}\frac{\dot r^2}{756}
\left[1273\nu^2-907\nu+131+(48-144\nu)\bar\gamma_{AB}\right]
+\frac{3545\nu^2-1835\nu+253}{504}v^4,\label{eq:CM-I2-A2}\\
\mathcal B_2={}&-\frac{G_{AB}m}{189r}
\left[985\nu^2+335\nu-742+(504\nu-420)\bar\gamma_{AB}\right]
+\frac{733\nu^2-337\nu+41}{126}v^2
+\frac{5(5\nu^2-5\nu+1)}{63}\dot r^2,\label{eq:CM-I2-B2}\\
\mathcal C_2={}&\frac{G_{AB}m}{378r}
\left[1463\nu^2+4057\nu-1085+(2844\nu-696)\bar\gamma_{AB}\right]
-\frac{2(209\nu^2-101\nu+13)}{63}v^2.\label{eq:CM-I2-C2}
\end{align}
\end{widetext}

Notice that these coefficients depend on the separation and
relative velocity in the coordinates used here,
the symmetric mass ratio, the relative mass difference,
and the scalar responses. The expressions apply to
generic orbits; no circular-orbit restriction has
been imposed.

The remaining moments can be organized in the same
way by expanding them in a finite basis of STF tensors
built from $\vn$ and $\vv$. For the current moments,
we also introduce the vector
$\boldsymbol\ell=\vn\times\vv$.
The two bases are
\begin{align}
\mathcal N_{\ell j}^{L}
&=\STF_L
\left[
n^{i_1}\cdots n^{i_{\ell-j}}
v^{i_{\ell-j+1}}\cdots v^{i_\ell}
\right],
\nonumber\\
\mathcal K_{\ell j}^{L}
&=\STF_L
\left[
n^{i_1}\cdots n^{i_{\ell-1-j}}
v^{i_{\ell-j}}\cdots v^{i_{\ell-1}}
\ell^{i_\ell}
\right].
\label{eq:CM-basis}
\end{align}
Here $j$ counts the explicit factors of $\vv$ outside
$\boldsymbol\ell$: it ranges from $0$ to $\ell$ in
$\mathcal N_{\ell j}^{L}$ and from $0$ to $\ell-1$
in $\mathcal K_{\ell j}^{L}$.
Empty products are understood to equal one.

With these definitions, the integer-PN parts of the moments take the common form
\begin{align}
I_L
&=\mu r^\ell\sum_{n,j}
c^{-2n}\mathcal I_{\ell,n,j}
\dot r^{\eta_j}\mathcal N_{\ell j}^{L},
\nonumber\\
J_L
&=\mu r^\ell\sum_{n,j}
c^{-2n}\mathcal J_{\ell,n,j}
\dot r^{\eta_j}\mathcal K_{\ell j}^{L},
\nonumber\\
Q_L^{\rm int}
&=\mu r^\ell\sum_{n,j}
c^{-2n}\mathcal Q_{\ell,n,j}
\dot r^{\eta_j}\mathcal N_{\ell j}^{L}
\quad(\ell\ge1),
\nonumber\\
Q
&=m\sum_{n=0}^2c^{-2n}\mathcal Q_{0,n,0},
\label{eq:CM-all-moments}
\end{align}
where $\eta_j=0$ for even $j$ and $\eta_j=1$ for
odd $j$. Here $n$ runs only over the integer-PN
orders retained for each moment in Table~\ref{tab:moments},
and $Q_L^{\rm int}$ denotes the corresponding scalar part.
All nonzero coefficient functions
$\mathcal I_{\ell,n,j}$, $\mathcal J_{\ell,n,j}$,
and $\mathcal Q_{\ell,n,j}$ at these orders are
provided in the ancillary file, with the reconstruction conventions in Appendix~\ref{app:CM-other}.

The scalar dipole also receives the 1.5PN term induced
by the radiation-reaction shift in Eq.~\eqref{eq:CM-shift-RR}:
\begin{align}
\delta Q_i^{\rm CM}
&=Q_{[0]}\delta x_{i,\mathrm{RR}}
=\frac{Gm\mu(\alpha_A-\alpha_B)\bar\alpha^2}{3c^3}v_i,
\label{eq:Q1-CM-RR}\\
Q_i&=Q_i^{\rm int}+\delta Q_i^{\rm CM}+\order(c^{-5}).
\label{eq:Q1-CM-complete}
\end{align}
This term reproduces the odd 1.5PN contribution to the
source dipole in Ref.~\cite{BBT2022}. It is written
separately because its velocity dependence does not
have the time-reversal parity of the integer-PN basis
in Eq.~\eqref{eq:CM-all-moments}.

The leading expressions illustrate how the mass
asymmetry and scalar charges enter:
\begin{align}
I_{ijk,[0]}
&=-\mu\Delta r^3n^{\langle ijk\rangle},
\nonumber\\
J_{ij,[0]}
&=-\mu\Delta r^2n^{\langle i}\ell^{j\rangle},
\nonumber\\
Q_{i,[0]}
&=\mu(\alpha_A-\alpha_B)r n^i,
\nonumber\\
Q_{ij,[0]}
&=\mu
\left(
\alpha_A\frac{m_B}{m}
+\alpha_B\frac{m_A}{m}
\right)
r^2n^{\langle ij\rangle}.
\label{eq:CM-leading}
\end{align}
In particular, equal scalar charges eliminate
the leading scalar dipole, while the scalar
quadrupole generally remains nonzero.

To obtain the instantaneous power, we differentiate the
moments and apply the conservative equations of motion through 2PN and
the leading reaction force in Eq.~\eqref{eq:relative-reaction}.
The integer-PN contributions retain the same reductions:
2PN acceleration corrections act on leading moments,
and 1PN corrections act on their 1PN terms. The additional
1.5PN terms arise from the reaction force in derivatives
of the leading moments and from differentiating
Eq.~\eqref{eq:Q1-CM-RR} with the Newtonian equations.
Derivatives of the reaction force are included whenever
higher derivatives of a moment require them. All reductions
are performed for generic motion before imposing circularity.

\section{Instantaneous energy flux}\label{sec:flux}

The source moments of Sec.~\ref{sec:results} determine
the linear radiation amplitudes. Their squares, integrated
over emission directions, give the outgoing powers
$P_T^{\rm lin}$ and $P_S^{\rm lin}$. We first state these
linear-emission expressions, then add the nonlinear
propagation contribution needed for the instantaneous
tensor flux. With the normalization of
Eq.~\eqref{eq:rad-action}, the linear powers are

\begin{widetext}
\begin{align}
P_T^{\rm lin}
={}&G\sum_{\ell\ge2}
\left[
\frac{(\ell+1)(\ell+2)}
{c^{2\ell+1}\ell(\ell-1)\ell!(2\ell+1)!!}
I_L^{(\ell+1)}I_L^{(\ell+1)}
+\frac{4\ell(\ell+2)}
{c^{2\ell+3}(\ell-1)(\ell+1)!(2\ell+1)!!}
J_L^{(\ell+1)}J_L^{(\ell+1)}
\right],
\label{eq:tensor-power-general}\\
P_S^{\rm lin}
={}&G\sum_{\ell\ge0}
\frac{Q_L^{(\ell+1)}Q_L^{(\ell+1)}}
{c^{2\ell+1}\ell!(2\ell+1)!!}.
\label{eq:scalar-power-general}
\end{align}
\end{widetext}
Here, a parenthesized superscript denotes the number of
time derivatives. In the scalar expression, the
overall factor $G$ follows from the canonical
coupling strength $\kappa/\sqrt8$, whose square
divided by $4\pi$ equals $G$; the remaining
coefficients arise from the STF angular integrals.

The required truncations follow from the moments
listed in Table~\ref{tab:moments}.
For tensor radiation, we retain the mass quadrupole
through 2PN, the mass octupole and current quadrupole
through 1PN, and the leading mass hexadecapole and
current octupole. For scalar radiation, we retain
the dipole through 2PN, the quadrupole through 1PN,
and the leading octupole. The monopole contributes
through the square of its differentiated 1PN term
and the interference between its differentiated
1PN and 2PN terms. All time derivatives and products
are expanded consistently to the required PN order,
including the acceleration corrections described
in Sec.~\ref{sec:results}.

\subsection{Nonlinear propagation and instantaneous power}
\label{sec:nonlinear-propagation}

The radiation measured far from the binary is described
by radiative moments. Let $U_{ij}$ denote the tensor
radiative quadrupole, so that its contribution to the
power is $G U_{ij}^{(1)}U_{ij}^{(1)}/(5c^5)$.
At linear order $U_{ij}=I_{ij}^{(2)}$. Scalar radiation
also sources the metric through its stress-energy tensor,
giving a nonlinear correction to this relation.
We adopt the gauge-completed relation of
Ref.~\cite{BBT2022}, Eq.~(3.28), translated to our
Einstein-frame moments:
\begin{widetext}
\begin{align}
U_{ij}(u)&=I_{ij}^{(2)}(u)+U_{ij}^{\rm tail}(u)
+\delta U_{ij}^{\varphi\varphi}(u)+\order(c^{-5}),
\label{eq:radiative-quadrupole}\\
\delta U_{ij}^{\varphi\varphi}(u)
&=\frac{G}{3c^3}\left[
\int_{-\infty}^{u}\dd u'\,
Q_{\langle i}^{(2)}(u')Q_{j\rangle}^{(2)}(u')
-Q_{\langle i}Q_{j\rangle}^{(3)}
-Q_{\langle i}^{(1)}Q_{j\rangle}^{(2)}
-\frac12 Q_{[0]}Q_{ij}^{(3)}\right].
\label{eq:scalar-nonlinear-map}
\end{align}
\end{widetext}
Here $u$ is the asymptotic retarded time,
$U_{ij}^{\rm tail}$ is the tail contribution,
and $\delta U_{ij}^{\varphi\varphi}$ denotes the correction
quadratic in scalar fields. Moments without an explicit
argument in Eq.~\eqref{eq:scalar-nonlinear-map} are
evaluated at $u$, and only their leading expressions are
needed on its right-hand side. The constant scalar
monopole $Q_{[0]}$ contributes through its coupling to
the scalar quadrupole even though it does not radiate
in the linear-emission formula.

The integral is the memory induced by scalar radiation;
the other three terms are its local propagation partners.
Their coefficients are taken from Ref.~\cite{BBT2022}.
An independent EFT derivation of this nonlinear relation
lies outside the present source-matching calculation
and is left for future work.
Although memory depends on the past history of the source,
its derivative contributes locally to the power. On
differentiating the full relation, we obtain 
\begin{align}
\delta U_{ij}^{(1),\varphi\varphi}
=-\frac{G}{3c^3}\left[
Q_{\langle i}Q_{j\rangle}^{(4)}
+2Q_{\langle i}^{(1)}Q_{j\rangle}^{(3)}
+\frac12Q_{[0]}Q_{ij}^{(4)}\right].
\label{eq:scalar-nonlinear-news}
\end{align}
Its interference with the leading tensor wave gives
\begin{equation}
\delta P_T^{\rm prop}
=\frac{2G}{5c^5}I_{ij,[0]}^{(3)}
\delta U_{ij}^{(1),\varphi\varphi},
\label{eq:nonlinear-tensor-power}
\end{equation}
where the label $\mathrm{prop}$ identifies propagation.
This term is of order $c^{-8}$, or 1.5PN relative to
the leading tensor power, and must accompany the
reaction-force insertion in the linear-emission formula.
In the scalar channel, no analogous local nonlinear
radiative-moment correction contributes through $c^{-7}$;
its additional local term at $c^{-6}$ comes from the
reaction force and the dissipative center-of-mass shift.

We follow the separation between instantaneous and tail
contributions adopted in Ref.~\cite{BBT2022}.
The scalar-induced memory becomes local upon differentiation
and therefore contributes to the instantaneous power.
The tail terms retain integrals over the past source evolution
and are not included in the fluxes presented below.
Following the same convention, we keep the constant pieces
of the tail kernels within the tail contribution, even when
they can be expressed locally. The complete outgoing flux
requires adding the tensor and scalar tail contributions.

\subsection{Fluxes for generic orbits}

The integer-PN parts follow from the conservative
reduction of the center-of-mass moments and retain
the finite polynomial form

\begin{widetext}
\begin{align}
P_T^{\rm int}
&=\frac{G^3m^4\nu^2}{c^5r^4}
\sum_{n=0}^2\frac1{c^{2n}}
\sum_{i+j+k=n+1}
F^T_{n;ijk}
\left(\frac{Gm}{r}\right)^i
(v^2)^j\dot r^{2k},
\label{eq:flux-CM-T}\\
P_S^{\rm int}
&=\frac{G^3m^4\nu^2}{c^3r^4}
\sum_{n=0}^2\frac1{c^{2n}}
\sum_{i+j+k=n}
F^S_{n;ijk}
\left(\frac{Gm}{r}\right)^i
(v^2)^j\dot r^{2k}.
\label{eq:flux-CM-S}
\end{align}
\end{widetext}
Here $n$ labels the PN order relative to the leading
contribution in each channel, and $i$, $j$, and $k$
are nonnegative integers specifying the powers of
the orbital variables. The coefficient functions
$F^T_{n;ijk}$ and $F^S_{n;ijk}$ depend only on
$\nu$, $\Delta$, and the scalar responses.
Their nonzero values are supplied in the ancillary Mathematica notebook,
with the normalization specified in Appendix~\ref{app:flux-coefficients}.

The remaining local contributions are obtained by
inserting Eq.~\eqref{eq:relative-reaction} into the
leading moment derivatives, adding the scalar center-of-mass term
in Eq.~\eqref{eq:Q1-CM-RR}, and evaluating
Eq.~\eqref{eq:nonlinear-tensor-power}. Their sum gives
\begin{widetext}
\begin{align}
\Delta P_T^{\rm inst}
&=\frac{8G^4m^5\nu^2(1+\alpha_A\alpha_B)^2\dot r}{45c^8r^5}
\left\{\left[\alpha_A\alpha_B+3\nu(\alpha_A-\alpha_B)^2\right]
\frac{Gm(1+\alpha_A\alpha_B)}{r}
+\alpha_A\alpha_B(35v^2-33\dot r^2)\right\},
\label{eq:flux-half-T}\\
\Delta P_S^{\rm inst}
&=-\frac{4G^4m^5\nu^2(1+\alpha_A\alpha_B)^2
(\alpha_A-\alpha_B)^2\dot r}{9c^6r^5}
\frac{m_A\alpha_A^2+m_B\alpha_B^2}{m}.
\label{eq:flux-half-S}
\end{align}
\end{widetext}
Both are at relative 1.5PN in their respective channels.
The scalar result combines terms proportional to
$\nu(\alpha_A-\alpha_B)^2$ from the relative reaction
force and to $\bar\alpha^2$ from the center-of-mass displacement,
using
\begin{equation}
\bar\alpha^2+\nu(\alpha_A-\alpha_B)^2
=\frac{m_A\alpha_A^2+m_B\alpha_B^2}{m}.
\label{eq:charge-weight-identity}
\end{equation}
The instantaneous powers at the stated accuracies are
\begin{equation}
P_T^{\rm inst}=P_T^{\rm int}+\Delta P_T^{\rm inst},
\qquad
P_S^{\rm inst}=P_S^{\rm int}+\Delta P_S^{\rm inst}.
\label{eq:complete-instantaneous-power}
\end{equation}
The additional terms are odd in $\dot r$, so they are
displayed separately from the even polynomials in
Eqs.~\eqref{eq:flux-CM-T} and \eqref{eq:flux-CM-S}.
No higher scalar-response derivatives enter these
additional terms;
the $\beta_A$ and $\beta'_A$ dependence remains in the
integer-PN coefficients.

Equal scalar charges remove the leading dipolar reaction
and Eq.~\eqref{eq:flux-half-S}, but do not generally
remove Eq.~\eqref{eq:flux-half-T} for noncircular motion:
the scalar monopole--quadrupole propagation contribution
can remain. Both additional terms vanish in the limit of
general relativity.

The leading terms display the different reference
orders of the two channels:
\begin{align}
P_{T,[0]}
&=\frac{8G^3m^4\nu^2(1+\alpha_A\alpha_B)^2}
{15c^5r^4}
(12v^2-11\dot r^2),
\label{eq:tensor-leading}\\
P_{S,[0]}
&=\frac{G^3m^4\nu^2(1+\alpha_A\alpha_B)^2}
{3c^3r^4}
(\alpha_A-\alpha_B)^2.
\label{eq:scalar-leading}
\end{align}
The scalar dipole flux begins one PN order before
the tensor quadrupole flux, as indicated by their
respective powers of $c$.

For a bound adiabatic inspiral, the orbit-averaged total
flux determines the secular decrease of the
binding energy. Before averaging, the balance also involves
the reversible near-zone contribution illustrated by
Eq.~\eqref{eq:dipole-energy-balance}. The local 1.5PN terms
have zero average on a fixed Newtonian ellipse, by their
odd parity in $\dot r$ with the orbital averaging measure.
A complete secular prediction nevertheless requires the
tail contributions at the  desired PN order.

\subsection{Circular orbits and comparisons}

The circular limit provides a useful connection
to inspiral modeling and a direct setting for
comparison with established results.
We obtain it by imposing the
conservative circular equations after differentiating
the moments for generic motion. The circular reduction
and the conversion to the frequency parameter are
detailed in Appendix~\ref{app:circular}.
For an attractive weak-field
circular orbit, $G_{AB}>0$, and we introduce
the dimensionless frequency parameter
$x=(G_{AB}m\Omega/c^3)^{2/3}$, where $\Omega$
is the orbital angular frequency.
After the generic derivatives have been evaluated,
Eqs.~\eqref{eq:flux-half-T} and \eqref{eq:flux-half-S}
vanish on setting $\dot r=0$. The circular instantaneous
powers are therefore determined by the integer-PN terms.
For an adiabatic quasi-circular inspiral, the leading
dipolar radial drift is of order $c^{-3}$, which promotes
these local terms to $c^{-11}$ and $c^{-9}$ in the tensor
and scalar powers, respectively, beyond the retained
accuracies. Tail contributions can survive circularity
and remain excluded here. The instantaneous fluxes are

\begin{align}
P_{T,\rm circ}^{\rm inst}
&=\frac{c^5\nu^2x^5}
{G(1+\alpha_A\alpha_B)^2}
\left[
\frac{32}{5}+f^T_1x+f^T_2x^2
\right],
\label{eq:ST-circular-T}\\
P_{S,\rm circ}^{\rm inst}
&=\frac{c^5\nu^2x^4}
{G(1+\alpha_A\alpha_B)^2}
\left[
\frac{(\alpha_A-\alpha_B)^2}{3}
+f^S_1x+f^S_2x^2
\right].
\label{eq:ST-circular-S}
\end{align}
The dimensionless coefficients $f^T_n$ and $f^S_n$
depend on the mass ratios and scalar responses.
Appendix~\ref{app:circular} expresses them
algebraically in terms of the generic flux
coefficients and the 2PN Kepler relation.

Setting both bodies' scalar responses to zero
recovers the limit of general relativity.
The scalar flux vanishes, and the tensor
coefficients reduce to
\begin{align}
\left.f^T_1\right|_{\alpha=\beta=\beta'=0}
&=\frac{32}{5}
\left(-\frac{1247}{336}-\frac{35}{12}\nu\right),
\nonumber\\
\left.f^T_2\right|_{\alpha=\beta=\beta'=0}
&=\frac{32}{5}
\left(
-\frac{44711}{9072}
+\frac{9271}{504}\nu
+\frac{65}{18}\nu^2
\right).
\label{eq:GR-circular}
\end{align}
Together with Eq.~\eqref{eq:ST-circular-T},
these coefficients reproduce the instantaneous
circular flux of general relativity through 2PN
for arbitrary mass
ratio~\cite{PetersMathews1963,BlanchetDamourIyer1995,WillWiseman1996,GopakumarIyer1997,Blanchet2014,Leibovich2020,Amalberti2024flux}.
The 1.5PN tail contribution belongs to the
hereditary sector and is therefore absent from
this instantaneous result. 
The separation between instantaneous and tail terms
also enters higher-order flux
calculations~\cite{Blanchet2023flux}.

For generic scalar--tensor orbits, the tensor flux
through relative 1PN agrees with Ref.~\cite{Lang2015}.
After translating the scalar--tensor parameters and
applying the coordinate transformation in
Eq.~\eqref{eq:gauge-shift}, the scalar flux coefficients
at relative 0PN, 1PN, and 2PN agree with
Eqs.~(C2)--(C3) of Ref.~\cite{BBT2022}. In particular,
this comparison confirms all six independent
coefficients at 2PN relative to the leading dipole.
These comparisons concern the integer-PN terms.
The scalar term in Eq.~\eqref{eq:flux-half-S} also
agrees with the relative 1.5PN contribution in those
expressions, completing the comparison of the scalar
instantaneous flux through dipole-relative 2PN.
The three orbital coefficients in
Eq.~\eqref{eq:flux-half-T} reproduce the local relative
1.5PN tensor term in Eq.~(4.11) of Ref.~\cite{BBT2022}.
This tensor comparison tests the reaction-force reduction
and multipolar contractions using the adopted nonlinear
memory relation; the relation itself is an input from that
reference.

In the general relativity limit, the 2PN coefficients
of the center-of-mass mass quadrupole agree with
Eq.~(6.16) of Ref.~\cite{Leibovich2020}. The instantaneous
energy flux also agrees, for generic motion, with
the 2PN result of Ref.~\cite{GopakumarIyer1997} after
the same coordinate transformation. Its circular
specialization reproduces the coefficients in
Eq.~\eqref{eq:GR-circular} for arbitrary symmetric
mass ratio.
\section{Conclusions}\label{sec:conclusions}

We have developed an explicit EFT construction of the
tensor and scalar radiation sources of nonspinning
compact binaries in massless scalar--tensor theories,
carrying out their matching onto the source multipoles.
Combining these moments with the conservative dynamics,
the leading dipolar radiation reaction, and the nonlinear
radiative-moment relation adopted from Ref.~\cite{BBT2022},
we have obtained the instantaneous tensor flux through
quadrupole-relative 2PN and the instantaneous scalar flux
through dipole-relative 2PN. The causal EFT calculation
also determines the dipolar reaction force and its
Schott and center-of-mass contributions.
These results retain generic scalar
charges and nonlinear compact-body responses, and are
given for arbitrary orbital motion together with their
circular specialization in terms of the gauge-invariant orbital frequency.

The calculation connects the compact-body description,
orbital dynamics, and radiation through a common gauge
and regularization prescription. Organizing the interactions
into eleven topology classes allows the composite sources
and their multipole projections to be assembled systematically.
The agreement between the source-derived conserved quantities
and the Noether charges, together with the cancellation of
dimensional-regularization poles in the retained multipole
projections, provides internal checks of this construction.
Independent comparisons establish agreement with the
generic scalar--tensor tensor flux through relative
1PN~\cite{Lang2015} and confirm the scalar instantaneous
flux through dipole-relative 2PN~\cite{BBT2022}, including
its local 1.5PN term, after translating the parameters
and coordinates. The local tensor contribution at
relative 1.5PN also agrees with Ref.~\cite{BBT2022}
when the adopted nonlinear memory relation is combined with
our source moments and reaction force.

In the general relativity limit, the center-of-mass
mass-quadrupole coefficients agree with those presented in 
Ref.~\cite{Leibovich2020}, while the instantaneous flux
reproduces the generic-orbit result through
2PN~\cite{GopakumarIyer1997} and its circular limit
for arbitrary mass ratio.

The nonlinear tensor propagation relation is a published
input to this calculation. Its independent derivation
within EFT is beyond the present scope and will be
addressed in future work. The instantaneous result includes
the derivative of scalar-induced memory and its local
partners, while tail integrals remain separate.

A natural extension is to bring both radiation channels
to a common quadrupole-relative 2PN accuracy, which
requires the scalar channel through dipole-relative 3PN.
Its integer-PN source calculation must determine
the scalar monopole and dipole through 3PN, the quadrupole
through 2PN, the octupole through 1PN, and the leading
hexadecapole. This entails extending the scalar source
to order $v^6$ and using the conservative dynamics
through 3PN in compatible coordinates. At that order,
the monopole flux also receives the square of its
2PN derivative, alongside the interference between
its 1PN and 3PN contributions.
Completing that flux also requires the dissipative
and nonlinear propagation contributions at the same
accuracy, which, in particular, will include higher-order 
tail and memory effects.

This work establishes an explicit EFT description
of radiation generation in massless scalar--tensor
theories, connecting compact-body interactions
to tensor and scalar multipoles through matching.
The resulting flux coefficients independently
confirm established scalar--tensor results using
a different calculational framework. Beyond this
confirmation, the explicit sources and multipoles
organize higher-order calculations through scale
separation and PN power counting. 
Together with the leading causal reaction derived here,
they provide a starting point for higher-order radiation
reaction. Combining these ingredients with hereditary
propagation will connect the source matching to the
secular evolution of compact binaries within the same
EFT framework.

\begin{acknowledgments}
This work was partially supported by the Start-up Research Fund of Southeast University under Grant No. RF1028624160.
\end{acknowledgments}

\appendix
\onecolumngrid
\section{Conservative dynamics and Noether charges}\label{app:action}
\subsection{Lagrangian and acceleration coefficients}\label{app:acceleration}
We give the conservative action in the linear harmonic coordinates used
for the source matching. We first retain positions, velocities, and
accelerations as independent arguments of the generalized Lagrangian,
\begin{equation}
L=-mc^2+L_0+c^{-2}L_1+c^{-4}L_{2,\rm lin}+\order(c^{-6}).
\label{eq:appendix-L-expansion}
\end{equation}
The Newtonian and 1PN coefficients are

\begin{align}
L_0={}&\frac12(m_Av_A^2+m_Bv_B^2)+\frac{G_{AB}m_Am_B}{r},\label{eq:L0}\\
L_1={}&\frac18(m_Av_A^4+m_Bv_B^4)
+\frac{G_{AB}m_Am_B}{2r}\left[(3+2\bar\gamma_{AB})(v_A^2+v_B^2)
-(7+4\bar\gamma_{AB})\vv_A\cdot\vv_B
-\frac{(\vr\cdot\vv_A)(\vr\cdot\vv_B)}{r^2}\right]\nonumber\\
&-\frac{G_{AB}^2m_Am_B}{2r^2}\left[m_A(1+2\bar\beta_B)+m_B(1+2\bar\beta_A)\right].\label{eq:L1}
\end{align}

To obtain the 2PN coefficient, we follow the diagrammatic calculation
of Ref.~\cite{Almeida2024}, accounting explicitly for the change in
gauge fixing. The nonlinear harmonic gauge-fixing action has the same
quadratic part as the linear harmonic gauge-fixing action, but contains
additional terms cubic and higher in the metric perturbation. We
recalculated all diagrams of Ref.~\cite{Almeida2024} involving these
additional vertices at the required order and subtracted their
contributions from the harmonic-gauge effective action. Through the
orders considered here, this subtraction affects only the 2PN
coefficient; $L_0$ and $L_1$ remain unchanged. Up to a total time
derivative, the result is
\begin{align}
L_{2,\rm lin}&=L_{2,\rm harm}+\delta L_2,\nonumber\\
\delta L_2&=\frac{2G^2m_Am_Bm}{r^2}
\left(v^2-2\dot r^2-\frac{G_{AB}m}{r}\right).
\label{eq:appendix-gauge-correction}
\end{align}
Here $L_{2,\rm harm}$ is the 2PN coefficient obtained in
Ref.~\cite{Almeida2024}. This form makes the result of the explicit
diagrammatic subtraction transparent without repeating the full
harmonic-gauge expression. The complete algebraic $L_{2,\rm lin}$ is
supplied in the ancillary notebook. Its acceleration-dependent terms
must be varied before substituting any equations of motion.

For clarity, write the Euler--Lagrange equations in the form
\begin{equation}
m_A\va_A=\mathbf F_{A,0}+c^{-2}\mathbf F_{A,1}
+c^{-4}\mathbf F_{A,2},\qquad
\mathbf F_{A,n}=\frac{\partial L_n}{\partial\vx_A}
-\frac{\dd}{\dd t}\frac{\partial L_n}{\partial\vv_A}
+\frac{\dd^2}{\dd t^2}\frac{\partial L_n}{\partial\va_A}
\quad(n=1,2),
\label{eq:appendix-force-reduction}
\end{equation}
where $L_2=L_{2,\rm lin}$ and
$\mathbf F_{A,0}=-G_{AB}m_Am_B\vr/r^3$.
The 1PN acceleration is obtained by evaluating $\mathbf F_{A,1}$
with Newtonian accelerations. At 2PN, we also retain the change in
$\mathbf F_{A,1}$ induced by the 1PN acceleration, while reducing
$\mathbf F_{A,2}$ and all of its higher derivatives with Newtonian
motion. This iteration yields the coefficients in
Eq.~\eqref{eq:acceleration}.

Their presentation simplifies substantially when the scalar responses
are collected into the parameters of Eqs.~\eqref{eq:ST-parameters1}
and \eqref{eq:ST-parameters2}. We also use
$\vn\cdot\vv_A=\dot r+\vn\cdot\vv_B$ to combine repeated radial
velocity terms. The resulting expressions are grouped by powers of
$G_{AB}$:
\begin{align}
\mathscr A_{A,1}={}&\frac{G_{AB}^{2} m_A m_B}{r^{4}}\Bigl[2 \bar\beta_B + 2 \bar\gamma_{AB} + 5\Bigr]\nonumber\\ &+\frac{G_{AB}^{2} m_B^{2}}{r^{4}}\Bigl[2 \bar\beta_A + 2 \bar\gamma_{AB} + 4\Bigr]\nonumber\\ &+\frac{G_{AB} m_B}{r^{3}}\Bigl[\frac{3 (\vn\cdot\vv_B)^{2}}{2} + v_A^{2} \left(- \bar\gamma_{AB} - 1\right) + v_B^{2} \left(- \bar\gamma_{AB} - 2\right) + 2 (\vv_A\cdot\vv_B) \left(\bar\gamma_{AB} + 2\right)\Bigr],\label{eq:acc-a1}
\end{align}

\begin{align}
\mathscr B_{A,1}={}&\frac{G_{AB} m_B}{r^{2}}\Bigl[2 \dot r \left(\bar\gamma_{AB} + 2\right) + (\vn\cdot\vv_B)\Bigr],\label{eq:acc-b1}
\end{align}

\begin{align}
\mathscr A_{A,2}={}&\frac{G_{AB}^{3} m_A^{2} m_B}{r^{5}}\Bigl[- \delta_A -  4 \bar\beta_B \left(\bar\gamma_{AB} + 3\right) -  9 \bar\gamma_{AB} -  \frac{7 \bar\gamma_{AB}^{2}}{4} -  \frac{\epsilon_B}{2} -  \frac{49}{4}\Bigr]\nonumber\\ &+\frac{G_{AB}^{3} m_A m_B^{2}}{r^{5}}\Bigl[- 20 \bar\gamma_{AB} + \bar\beta_A \left(- 4 \bar\gamma_{AB} - 15\right) -  4 \bar\beta_B \left(\bar\gamma_{AB} + 3\right) -  3 \zeta_{AB} -  3 \bar\gamma_{AB}^{2} -  \frac{61}{2}\Bigr]\nonumber\\ &+\frac{G_{AB}^{3} m_B^{3}}{r^{5}}\Bigl[- \delta_B -  4 \bar\beta_A \left(\bar\gamma_{AB} + 2\right) -  7 \bar\gamma_{AB} -  \frac{7 \bar\gamma_{AB}^{2}}{4} -  \frac{\epsilon_A}{2} - 7\Bigr]\nonumber\\ &+\frac{G_{AB}^{2} m_A m_B}{r^{4}}\Bigl[\dot r^{2} \left(- 4 \bar\beta_B + 2 \delta_A - \frac{7 \bar\gamma_{AB}^{2}}{2} - 3 \bar\gamma_{AB} + \frac{7}{2}\right) + (\vn\cdot\vv_B)^{2} \left(- 4 \bar\beta_B - 4 \bar\gamma_{AB} - 11\right)\nonumber\\ &\hspace{12mm}- \frac{v_A^{2} \left(4 \bar\beta_B - 4 \bar\gamma_{AB}^{2} - 8 \bar\gamma_{AB} - 1\right)}{4} + \frac{v_B^{2} \left(4 \bar\beta_B + 4 \bar\gamma_{AB}^{2} + 16 \bar\gamma_{AB} + 21\right)}{4}\nonumber\\ &\hspace{12mm}- \frac{(\vv_A\cdot\vv_B) \left(4 \bar\beta_B + 4 \bar\gamma_{AB}^{2} + 16 \bar\gamma_{AB} + 21\right)}{2}\Bigr]\nonumber\\ &+\frac{G_{AB}^{2} m_B^{2}}{r^{4}}\Bigl[\dot r^{2} \left(2 \delta_B - \frac{7 \bar\gamma_{AB}^{2}}{2} - 14 \bar\gamma_{AB} - 14\right) + (\vn\cdot\vv_B)^{2} \left(- 4 \bar\beta_A - 4 \bar\gamma_{AB} - 8\right)\nonumber\\ &\hspace{12mm}+v_A^{2} \left(- 2 \bar\beta_A + \bar\gamma_{AB}^{2} + 4 \bar\gamma_{AB} + 4\right) + v_B^{2} \left(\bar\gamma_{AB} + 2\right) \left(\bar\gamma_{AB} + 4\right)\nonumber\\ &\hspace{12mm}- 2 (\vv_A\cdot\vv_B) \left(\bar\gamma_{AB} + 2\right) \left(\bar\gamma_{AB} + 4\right)\Bigr]\nonumber\\ &+\frac{G_{AB} m_B}{r^{3}}\Bigl[- \frac{15 (\vn\cdot\vv_B)^{4}}{8} + \frac{3 (\vn\cdot\vv_B)^{2} v_A^{2} \left(\bar\gamma_{AB} + 1\right)}{2} + \frac{3 (\vn\cdot\vv_B)^{2} v_B^{2} \left(\bar\gamma_{AB} + 3\right)}{2}\nonumber\\ &\hspace{12mm}- 3 (\vv_A\cdot\vv_B) (\vn\cdot\vv_B)^{2} \left(\bar\gamma_{AB} + 2\right) + \left(v_B^{2}\right)^{2} \left(- \bar\gamma_{AB} - 2\right) + 2 (\vv_A\cdot\vv_B) v_B^{2} \left(\bar\gamma_{AB} + 2\right)\nonumber\\ &\hspace{12mm}+(\vv_A\cdot\vv_B)^{2} \left(- \bar\gamma_{AB} - 2\right)\Bigr],\label{eq:acc-a2}
\end{align}

\begin{align}
\mathscr B_{A,2}={}&\frac{G_{AB}^{2} m_A m_B}{r^{3}}\Bigl[\frac{\dot r \left(8 \bar\beta_B - 8 \delta_A + 6 \bar\gamma_{AB}^{2} - 8 \bar\gamma_{AB} - 31\right)}{4} -  2 (\vn\cdot\vv_B) \left(\bar\beta_B + 1\right)\Bigr]\nonumber\\ &+\frac{G_{AB}^{2} m_B^{2}}{r^{3}}\Bigl[- \frac{\dot r \left(4 \delta_B - 3 \bar\gamma_{AB}^{2} - 12 \bar\gamma_{AB} - 12\right)}{2} -  2 (\vn\cdot\vv_B) \left(\bar\beta_A + \bar\gamma_{AB} + 2\right)\Bigr]\nonumber\\ &+\frac{G_{AB} m_B}{r^{2}}\Bigl[- 3 \dot r (\vn\cdot\vv_B)^{2} \left(\bar\gamma_{AB} + 2\right) + 2 \dot r v_B^{2} \left(\bar\gamma_{AB} + 2\right) -  2 (\vv_A\cdot\vv_B) \dot r \left(\bar\gamma_{AB} + 2\right) -  \frac{3 (\vn\cdot\vv_B)^{3}}{2}\nonumber\\ &\hspace{12mm}+(\vn\cdot\vv_B) v_A^{2} \left(\bar\gamma_{AB} + 1\right) + (\vn\cdot\vv_B) v_B^{2} \left(- \bar\gamma_{AB} - 1\right)\Bigr].\label{eq:acc-b2}
\end{align}

These are algebraic rearrangements of the linear-harmonic acceleration,
including its gauge-dependent terms; collecting factors of $G_{AB}$
here does not replace $G$ by $G_{AB}$ in the action or in the gauge
displacement. Body $B$ follows by interchanging every body label,
with $\vr\to-\vr$, $\vn\to-\vn$, and $\vv\to-\vv$;
$\dot r=\vn\cdot\vv$ is unchanged.

In the limit of vanishing scalar responses, the difference from the conservative Lagrangian in Eq.~(B8) of Ref.~\cite{Leibovich2020} is the boundary term

\begin{equation}
\left.L_{2,\rm lin}\right|_{\alpha_A=\beta_A=\beta'_A=0}-L_{2,\rm Leibovich}
=\frac{\dd}{\dd t}\left[\frac{G^2m_Am_B}{r^2}\bigl(m_B\vr\cdot\vv_A-m_A\vr\cdot\vv_B\bigr)\right].
\label{eq:GR-boundary}
\end{equation}

The zero-response condition applies to both bodies. This comparison establishes the conservative relation in the stated coordinates. 

\subsection{Noether construction and explicit charges}\label{app:Noether}
For an arbitrary fixed-time worldline variation $\delta\vx_A$, integration by parts gives
\begin{equation}
\delta L=\sum_A\mathcal E_A\cdot\delta\vx_A
+\frac{\dd}{\dd t}\sum_A(\mathbf p_A\cdot\delta\vx_A+\mathbf q_A\cdot\delta\vv_A),
\label{eq:Noether-variation}
\end{equation}
where $\mathcal E_A$ is the Euler--Lagrange expression on the left-hand side of Eq.~\eqref{eq:EL}. For a symmetry with $\delta L=\dd F/\dd t$, the boundary expression minus $F$ is conserved. We apply it separately to each spacetime symmetry, keeping the
variation at fixed coordinate time.

\paragraph{Time translations.}
For a constant infinitesimal parameter $\varepsilon$, take
$\delta\vx_A=\varepsilon\vv_A$ and
$\delta\vv_A=\varepsilon\va_A$. Time independence of the
Lagrangian gives $\delta L=\varepsilon\dot L$, so that
$F=\varepsilon L$. Dividing the Noether charge by $\varepsilon$
gives
\begin{equation}
E=\sum_A(\mathbf p_A\cdot\vv_A+\mathbf q_A\cdot\va_A)-L.
\label{eq:appendix-energy-start}
\end{equation}
The constant $-mc^2$ in $L$ therefore contributes $+mc^2$ to $E$.

\paragraph{Spatial translations.}
For a constant vector $\boldsymbol\varepsilon$,
$\delta\vx_A=\boldsymbol\varepsilon$ and $\delta\vv_A=0$.
Since $L$ depends on positions through the separation,
$\delta L=0$. The coefficient of $\boldsymbol\varepsilon$ is
\begin{equation}
\mathbf P=\sum_A\mathbf p_A.
\label{eq:appendix-momentum-start}
\end{equation}

\paragraph{Rotations.}
For a constant infinitesimal rotation $\boldsymbol\theta$,
$\delta\vx_A=\boldsymbol\theta\times\vx_A$ and
$\delta\vv_A=\boldsymbol\theta\times\vv_A$.
Rotational invariance again gives $\delta L=0$.
Using the scalar triple-product identity in the boundary term yields
\begin{equation}
\mathbf L=\sum_A(\vx_A\times\mathbf p_A+\vv_A\times\mathbf q_A).
\label{eq:appendix-angular-start}
\end{equation}

\paragraph{Lorentz boosts.}

For an infinitesimal constant boost velocity $\mathbf b$, we use the fixed-time variation
\begin{equation}
\delta\vx_A=-\mathbf b\,t+\frac{\vv_A}{c^2}(\mathbf b\cdot\vx_A).
\label{eq:boost-variation}
\end{equation}
Its velocity variation is
\begin{equation}
\delta\vv_A=-\mathbf b+\frac1{c^2}
\left[\va_A(\mathbf b\cdot\vx_A)+\vv_A(\mathbf b\cdot\vv_A)\right],
\label{eq:boost-velocity-variation}
\end{equation}
and $\delta\va_A$ is obtained by differentiating once more.
Writing its Lagrangian variation as $\delta L=\dd(\mathbf b\cdot\mathbf Z)/\dd t$ defines the boost boundary vector $\mathbf Z$. The coefficient of $\mathbf b$ in the Noether charge is $\boldsymbol{\mathcal G}-t\mathbf P$, with
\begin{align}
\boldsymbol{\mathcal G}={}&\frac1{c^2}\sum_A\left[\vx_A(\mathbf p_A\cdot\vv_A+\mathbf q_A\cdot\va_A)
+\vv_A(\mathbf q_A\cdot\vv_A)\right]\nonumber\\
&-\sum_A\mathbf q_A-\mathbf Z.
\label{eq:boost-Noether}
\end{align}
To determine $\mathbf Z$ explicitly, split the boost into
$\delta=\delta_0+c^{-2}\delta_1$, as displayed in
Eqs.~\eqref{eq:boost-variation} and \eqref{eq:boost-velocity-variation},
and expand $\mathbf Z=\mathbf Z_0+c^{-2}\mathbf Z_1+c^{-4}\mathbf Z_2$.
At successive orders the starting equations are
\begin{align}
\frac{\dd}{\dd t}(\mathbf b\cdot\mathbf Z_0)&=\delta_0L_0,
&\frac{\dd}{\dd t}(\mathbf b\cdot\mathbf Z_1)&=\delta_0L_1+\delta_1L_0,
&\frac{\dd}{\dd t}(\mathbf b\cdot\mathbf Z_2)&=\delta_0L_{2,\rm lin}+\delta_1L_1.
\label{eq:boost-boundary-orders}
\end{align}
Here $\delta_nL$ means the full variation through positions, velocities,
and accelerations. Each expression is rearranged into a total time
derivative, to the required PN accuracy. For example, the first two give
\begin{equation}
\mathbf Z_0=-\sum_A m_A\vx_A,\qquad
\mathbf Z_1=\frac12\sum_A m_Av_A^2\vx_A
+\frac{G_{AB}m_Am_B}{2r}(\vx_A+\vx_B).
\label{eq:boost-boundary-low}
\end{equation}
These reproduce the Newtonian and 1PN parts of
$\boldsymbol{\mathcal G}$ below. At 2PN the same procedure, including
$\mathbf q_A$, gives the displayed coefficients $\mathscr B_A$ and
$\mathscr V_A$.

The charges are formed before replacing accelerations with their lower-order equations of motion. This order matters at 2PN because it retains the acceleration-conjugate terms in Eq.~\eqref{eq:Noether-charges}.

For the explicit 1PN energy, $\mathbf q_A$ first contributes beyond
the required order, and Eq.~\eqref{eq:appendix-energy-start} reduces to
\begin{equation}
E_0=\sum_A\vv_A\cdot\frac{\partial L_0}{\partial\vv_A}-L_0,
\qquad
E_1=\sum_A\vv_A\cdot\frac{\partial L_1}{\partial\vv_A}-L_1.
\label{eq:energy-coefficient-start}
\end{equation}
The kinetic quartic terms in $L_1$ acquire a factor three, its quadratic
velocity terms retain their coefficients, and its static terms change
sign. Thus the energy is $E=mc^2+E_0+c^{-2}E_1+\order(c^{-4})$, with

\begin{align}
E_0={}&\frac12(m_Av_A^2+m_Bv_B^2)-\frac{G_{AB}m_Am_B}r,\label{eq:energy0}\\
E_1={}&\frac38(m_Av_A^4+m_Bv_B^4)
+\frac{G_{AB}m_Am_B}{2r}\left[(3+2\bar\gamma_{AB})(v_A^2+v_B^2)
-(7+4\bar\gamma_{AB})\vv_A\cdot\vv_B
-\frac{(\vr\cdot\vv_A)(\vr\cdot\vv_B)}{r^2}\right]\nonumber\\
&+\frac{G_{AB}^2m_Am_B}{2r^2}\left[m_A(1+2\bar\beta_B)+m_B(1+2\bar\beta_A)\right].\label{eq:energy1}
\end{align}

For the momentum and angular momentum at 1PN, the starting expression is
$\mathbf p_A=m_A\vv_A+c^{-2}\partial L_1/\partial\vv_A+\order(c^{-4})$.
For example,
$\partial[(\vr\cdot\vv_A)(\vr\cdot\vv_B)]/\partial\vv_A
=(\vr\cdot\vv_B)\vr$.
Differentiating Eq.~\eqref{eq:L1} therefore gives

\begin{equation}
\mathbf p_A=m_A\vv_A+\frac1{c^2}\left\{\frac{m_Av_A^2}{2}\vv_A
+\frac{G_{AB}m_Am_B}{2r}\left[2(3+2\bar\gamma_{AB})\vv_A-(7+4\bar\gamma_{AB})\vv_B
-\frac{(\vr\cdot\vv_B)}{r^2}\vr\right]\right\}.
\label{eq:canonical-momentum}
\end{equation}
\begin{align}
\mathbf P={}&\mathbf p_A+\mathbf p_B+\order(c^{-4})\nonumber\\
={}&m_A\vv_A+m_B\vv_B+\frac1{2c^2}(m_Av_A^2\vv_A+m_Bv_B^2\vv_B)\nonumber\\ &
-\frac{G_{AB}m_Am_B}{2c^2r}
\left[\vv_A+\vv_B+\frac{\vr\cdot(\vv_A+\vv_B)}{r^2}\vr\right]+\order(c^{-4}),\label{eq:momentum}\\
\mathbf L={}&\vx_A\times\mathbf p_A+\vx_B\times\mathbf p_B+\order(c^{-4}).\label{eq:angular-momentum}
\end{align}

Equations~\eqref{eq:canonical-momentum} and \eqref{eq:angular-momentum} give the full angular momentum at the source order retained. The boost construction yields the mass dipole
\begin{align}
\boldsymbol{\mathcal G}={}&\sum_A m_A\vx_A\nonumber\\ &
+\frac1{2c^2}\sum_A\left[m_Av_A^2-\frac{G_{AB}m_Am_B}r\right]\vx_A\nonumber\\
&+\frac1{c^4}\left[\mathscr B_A\vx_A+\mathscr V_A\vv_A+(A\leftrightarrow B)\right]
+\order(c^{-6}),\label{eq:boost}
\end{align}
where the sum runs over the two bodies and the companion in each term is the other body. The 2PN coefficients are

\begin{align}
\mathscr B_A={}&\frac{G^{2} m_A^{2} m_B \left(\alpha_A^{2} \alpha_B^{2} + 2 \alpha_A^{2} \left(\alpha_B^2+\beta_B\right) - 2 \alpha_A \alpha_B - 5\right)}{4 r^{2}} - \frac{G^{2} m_A m_B^{2} \left(\alpha_A \alpha_B - 7\right) \left(\alpha_A \alpha_B + 1\right)}{4 r^{2}}\nonumber\\ &-  \frac{G (\vr\cdot\vv_A)^{2} m_A m_B \left(\alpha_A \alpha_B + 1\right)}{8 r^{3}} - \frac{G (\vr\cdot\vv_A) (\vr\cdot\vv_B) m_A m_B \left(\alpha_A \alpha_B + 1\right)}{4 r^{3}}\nonumber\\ &+ \frac{G (\vr\cdot\vv_B)^{2} m_A m_B \left(\alpha_A \alpha_B + 1\right)}{8 r^{3}} +\frac{G (\vv_A\cdot\vv_B) m_A m_B \left(\alpha_A \alpha_B - 7\right)}{4 r} -  \frac{G m_A m_B v_A^{2} \left(5 \alpha_A \alpha_B - 19\right)}{8 r}\nonumber\\ &+ \frac{G m_A m_B v_B^{2} \left(\alpha_A \alpha_B - 7\right)}{8 r} +\frac{3 m_A \left(v_A^{2}\right)^{2}}{8},\label{eq:boost-B}
\end{align}

\begin{align}
\mathscr V_A={}&\frac{G (\vr\cdot\vv_A) m_A m_B \left(\alpha_A \alpha_B - 7\right)}{4 r} + \frac{G (\vr\cdot\vv_B) m_A m_B \left(\alpha_A \alpha_B - 7\right)}{4 r}.\label{eq:boost-V}
\end{align}

These expressions satisfy $\dot{\boldsymbol{\mathcal G}}=\mathbf P$, including the 2PN momentum obtained from the generalized action. The energy and momenta agree with their source integrals as described in Sec.~\ref{sec:results}.

\subsection{Center-of-mass displacement}\label{app:CM-shift}
Write $\boldsymbol{\mathcal G}=\boldsymbol{\mathcal G}_0
+c^{-2}\boldsymbol{\mathcal G}_1+c^{-4}\boldsymbol{\mathcal G}_2$
and denote the Newtonian center-of-mass positions by
$\vx_A^N=(m_B/m)\vr$ and $\vx_B^N=-(m_A/m)\vr$.
Introduce a common displacement
$\delta\vx=c^{-2}\boldsymbol\eta_1+c^{-4}\boldsymbol\eta_2$.
The relative position is unchanged, while
$\boldsymbol{\mathcal G}_0=m\delta\vx$.
Expanding the condition $\boldsymbol{\mathcal G}=0$ gives
\begin{align}
\boldsymbol\eta_1&=-\frac1m
\left.\boldsymbol{\mathcal G}_1\right|_N,\nonumber\\
\boldsymbol\eta_2&=-\frac1m
\left[\left.\boldsymbol{\mathcal G}_2\right|_N
+\sum_{a=A,B}\left(
\boldsymbol\eta_1\cdot\frac{\partial}{\partial\vx_a}
+\dot{\boldsymbol\eta}_1\cdot\frac{\partial}{\partial\vv_a}
\right)\left.\boldsymbol{\mathcal G}_1\right|_N\right].
\label{eq:CM-recursion}
\end{align}
The derivatives act before the Newtonian center-of-mass substitution, indicated by
$|_N$, and $\dot{\boldsymbol\eta}_1$ is evaluated with Newtonian
motion. The first equation gives $\boldsymbol\eta_1=f_1\vr$, with
\begin{equation}
f_1=\frac{\nu\Delta}{2}\left(v^2-\frac{G_{AB}m}{r}\right).
\label{eq:CM-f1}
\end{equation}
Inserting this solution into the second equation and resolving the
result along $\vr$ and $r\vv$ gives
$\boldsymbol\eta_2=f_2\vr+b_2r\vv$, where

\begin{align}
f_2={}&- \frac{\Delta \alpha_A^{2} \alpha_B^{2} \left(\frac{Gm}{r}\right)^{2} \nu \left(2 \nu + 1\right)}{4} -  \frac{\alpha_A^{2} \left(\alpha_B^2+\beta_B\right) \left(\frac{Gm}{r}\right)^{2} \nu^{2}}{2} - \frac{\Delta \alpha_A \alpha_B \left(\frac{Gm}{r}\right)^{2} \nu \left(2 \nu - 3\right)}{2}\nonumber\\ &+ \frac{\alpha_B^{2} \left(\alpha_A^2+\beta_A\right) \left(\frac{Gm}{r}\right)^{2} \nu^{2}}{2} - \frac{\Delta \left(\frac{Gm}{r}\right)^{2} \nu \left(2 \nu - 7\right)}{4} + \frac{\Delta \alpha_A \alpha_B \left(\frac{Gm}{r}\right) \nu v^{2} \left(12 \nu - 5\right)}{8}\nonumber\\ &+\frac{\Delta \left(\frac{Gm}{r}\right) \nu v^{2} \left(12 \nu + 19\right)}{8} + \frac{\Delta \alpha_A \alpha_B \left(\frac{Gm}{r}\right) \dot r^{2} \nu \left(6 \nu - 1\right)}{8} +\frac{\Delta \left(\frac{Gm}{r}\right) \dot r^{2} \nu \left(6 \nu - 1\right)}{8} -  \frac{3 \Delta \nu v^4 \left(4 \nu - 1\right)}{8},\label{eq:CM-f2}
\end{align}

\begin{align}
b_2={}&\frac{\Delta \alpha_A \alpha_B \left(\frac{Gm}{r}\right) \dot r \nu}{4} -  \frac{7 \Delta \left(\frac{Gm}{r}\right) \dot r \nu}{4}.\label{eq:CM-b2}
\end{align}

The term proportional to $b_2$ is along the relative velocity; it is required for generic motion even though $b_2$ vanishes on a conservative circular orbit.

The dissipative displacement is obtained separately from the leading
scalar force. With $\mathbf Q=\sum_A m_A\alpha_A\vx_A$ and constant
$Q_{[0]}=\sum_A m_A\alpha_A$, its mechanical balances follow from
\begin{equation}
\dot E_{\rm mech}=\sum_A m_A\vv_A\cdot\va_{A,\rm RR}
=\frac{G}{3c^3}\dot{\mathbf Q}\cdot\mathbf Q^{(3)},\qquad
\dot{\mathbf P}_{\rm mech}=\sum_A m_A\va_{A,\rm RR}
=\frac{GQ_{[0]}}{3c^3}\mathbf Q^{(3)}.
\label{eq:appendix-mechanical-balances}
\end{equation}
Only the Newtonian charges are needed in these insertions: combining
1PN charge corrections with the 1.5PN force would contribute at 2.5PN.
The identity
$\dot{\mathbf Q}\cdot\mathbf Q^{(3)}
=\dd(\dot{\mathbf Q}\cdot\ddot{\mathbf Q})/\dd t-\ddot{\mathbf Q}^2$
gives the energy Schott term in Eq.~\eqref{eq:dipole-energy-balance}.
Similarly, integration of the momentum balance yields
$\mathbf P_{\rm Schott}=-GQ_{[0]}\ddot{\mathbf Q}/(3c^3)$.
To this order $\dot{\boldsymbol{\mathcal G}}_{\rm cons}=\mathbf P_{\rm mech}$;
hence the corrected mass dipole must satisfy
\begin{equation}
\delta\dot{\boldsymbol{\mathcal G}}=\mathbf P_{\rm Schott},
\qquad
\delta\boldsymbol{\mathcal G}=-\frac{GQ_{[0]}}{3c^3}\dot{\mathbf Q}.
\label{eq:appendix-dissipative-boost}
\end{equation}
The integration constant fixes the spatial origin. Setting the corrected
mass dipole to zero gives
$m\delta\vx_{\rm RR}=GQ_{[0]}\dot{\mathbf Q}/(3c^3)$,
which reproduces Eq.~\eqref{eq:CM-shift-RR} upon substituting the
Newtonian center-of-mass dipole. Thus this displacement supplements the conservative
recursion; it is not part of $f_1$, $f_2$, or $b_2$.

\section{Multipole coefficients and circular reduction}\label{app:coefficients}
The formulas below specify the tensor bases and normalizations needed
to reconstruct the source moments and instantaneous flux. Their long
coefficients are supplied as explicit expressions in the ancillary
Mathematica notebook \texttt{ST\_Radiative\_Coefficients.nb}.
Each section lists the corresponding quantities individually and
defines its notation. The notebook is self-contained and retains the
full dependence on $\alpha_A$, $\beta_A$, $\beta'_A$, and their
body-$B$ counterparts.

\subsection{Mass quadrupole before center-of-mass specialization}\label{app:general-I2}
The 2PN correction in Eq.~\eqref{eq:I2-general} is resolved onto the six
symmetric trace-free structures built from $\vx_A$, $\vx_B$, $\vv_A$,
and $\vv_B$, with coefficient functions
$C_{xx},C_{xy},C_{xv},C_{xw},C_{vv},C_{vw}$. These functions contain the
kinematic and potential-interaction contributions, including terms
induced by reducing lower-order accelerations. Their scalar arguments
are the masses, scalar responses, separation, and Euclidean products
of the position and velocity vectors. The notebook gives them as
\texttt{Cxx}, \texttt{Cxy}, \texttt{Cxv}, \texttt{Cxw}, \texttt{Cvv}, and
\texttt{Cvw}, respectively. 
The coefficients supplied in the ancillary notebook correspond
to the ordered-body terms displayed in Eq.~\eqref{eq:I2-general}.
The contribution denoted by $(A\leftrightarrow B)$ is obtained
by exchanging all body labels in both these coefficients
and their accompanying tensors, including
$\vr=\vx_A-\vx_B\to-\vr$.

\subsection{Other moments before center-of-mass specialization}\label{app:general-other}
For the scalar dipole at 2PN, mass octupole at 1PN, and scalar quadrupole at 1PN, we define the following ordered-body tensor expansion:

\begin{equation}
M_{L,[n]}=\sum_{a+b+c+d=\ell}D^{M,\ell}_{abcd}\,
\STF_L[\vx_A^{\otimes a}\vx_B^{\otimes b}\vv_A^{\otimes c}\vv_B^{\otimes d}]+(A\leftrightarrow B).
\label{eq:general-D-basis}
\end{equation}

Only the nonzero ordered-body coefficients supplied in the ancillary file are included in this sum. Here $(M,\ell,n)=(Q,1,2),(I,3,1),(Q,2,1)$, respectively; tensor products are symmetrized before trace subtraction. The scalar monopole is given directly. Together with Eqs.~\eqref{eq:leading-moments}, \eqref{eq:Q0-1PN}, and \eqref{eq:Q1-1PN}, these formulas reconstruct the retained scalar moments before specializing to the center-of-mass frame.
The notebook lists the full monopole coefficient $Q_{[2]}$ as
\texttt{Qmon2}. The functions \texttt{DQ1[a,b,c,d]},
\texttt{DI3[a,b,c,d]}, and \texttt{DQ2[a,b,c,d]} give
$D^{Q,1}_{abcd}$, $D^{I,3}_{abcd}$, and $D^{Q,2}_{abcd}$,
respectively. In particular,
\begin{align}
Q^i_{[2]}&=D^{Q,1}_{1000}x_A^i+D^{Q,1}_{0010}v_A^i+(A\leftrightarrow B),\nonumber\\
I^{ijk}_{[1]}&=\STF_{ijk}\left[
D^{I,3}_{3000}x_A^ix_A^jx_A^k+D^{I,3}_{2100}x_A^ix_A^jx_B^k
+D^{I,3}_{2010}x_A^ix_A^jv_A^k+D^{I,3}_{1020}x_A^iv_A^jv_A^k
\right]+(A\leftrightarrow B),\nonumber\\
Q^{ij}_{[1]}&=\STF_{ij}\left[
D^{Q,2}_{2000}x_A^ix_A^j+D^{Q,2}_{1100}x_A^ix_B^j
+D^{Q,2}_{1010}x_A^iv_A^j+D^{Q,2}_{0020}v_A^iv_A^j
\right]+(A\leftrightarrow B).
\label{eq:ancillary-general-reconstruction}
\end{align}
The interchange applies to coefficients as well as tensors, including
$\vr\to-\vr$. Mixed tensor structures that map to themselves already
carry the normalization required by the explicit interchange; no extra
factor of two is to be inserted. The scalar monopole coefficient is the full
two-body expression and requires no interchange.

\subsection{Current quadrupole and source operators}\label{app:current}
For the current quadrupole, let $(\mathbf u_0,\mathbf u_1,\mathbf u_2,\mathbf u_3)=(\vx_A,\vr,\vv_A,\vv_B)$. The complete first correction is
\begin{equation}
J^{ab}_{[1]}=\sum_{i=0}^3\sum_{j<k}E_{ijk}\,u_i^{\langle a}(\mathbf u_j\times\mathbf u_k)^{b\rangle}.
\label{eq:general-current}
\end{equation}
There is no additional body interchange in this formula. The nonzero
$E_{ijk}$ are listed as \texttt{Ecurrent[i,j,k]} in the notebook,
with $0\leq i\leq3$ and $0\leq j<k\leq3$.
The order of the cross product is fixed by $j<k$ and must be preserved.
The source operators yielding the additional tensor moments are, to the required orders,

\begin{align}
I^{ijk}&=\STF_{ijk}\int\dd^3x\,x^ix^jx^k\left[T^{00}+T^{ll}-\partial_t(x_aT^{0a})+\frac16\partial_t^2(x^2T^{00})\right],\label{eq:I3-operator}\\
J^{ij}&=\STF_{ij}\int\dd^3x\,\epsilon^{iab}x^jx^a\left[T^{0b}+\frac3{28}\partial_t^2(x^2T^{0b})-\frac14\partial_t(x_cT^{bc})\right].\label{eq:J2-operator}
\end{align}
Here, the symbol $\epsilon^{ijk}$ denotes the three-dimensional Levi-Civita tensor, with $\epsilon^{123}=1$.

The undifferentiated sources in $I^{ijk}$ are $T^{00}_{(v^0)}+T^{00}_{(v^2)}$ and $T^{ll}_{(v^2)}$; its differentiated terms use $T^{0a}_{(v)}$ and $T^{00}_{(v^0)}$. In $J^{ij}$ the undifferentiated source is $T^{0b}_{(v)}+T^{0b}_{(v^3)}$, while the differentiated sources are $T^{0b}_{(v)}$ and $T^{bc}_{(v^2)}$. The leading higher ranks follow from Eq.~\eqref{eq:leading-moments}.

\subsection{Center-of-mass multipoles}\label{app:CM-other}
The notebook lists the nonzero integer-PN coefficients in
Eq.~\eqref{eq:CM-all-moments} as \texttt{ICM[l,n,j]},
\texttt{JCM[l,n,j]}, and \texttt{QCM[l,n,j]}, where \texttt{l} denotes
the multipole rank $\ell$. The separate mass-quadrupole coefficients
shown in the main text are also supplied as \texttt{AI2[n]},
\texttt{BI2[n]}, and \texttt{CI2[n]}, for $n=1,2$.
The quantities $\nu$ and $\Delta$ obey $\Delta^2=1-4\nu$; the formulas are regular at equal masses and at equal scalar charges.
The mass and current moments carry the overall factor $\mu r^\ell$,
as do the scalar moments with $\ell\geq1$; the scalar monopole carries
$m$. For odd numbers $j$ of relative velocities in the tensor basis,
the extra factor $\dot r$ is written outside the coefficient in
Eq.~\eqref{eq:CM-all-moments}. The file follows precisely this convention.
For example,
\begin{equation}
Q_i^{\rm int}=\mu r\left[
\mathcal Q_{1,0,0}n_i+\frac{\mathcal Q_{1,1,0}n_i+
\mathcal Q_{1,1,1}\dot r\,v_i}{c^2}
+\frac{\mathcal Q_{1,2,0}n_i+\mathcal Q_{1,2,1}\dot r\,v_i}{c^4}
\right].\label{eq:ancillary-CM-dipole}
\end{equation}
The dissipative center-of-mass correction of Eq.~\eqref{eq:Q1-CM-RR} must be
added to this integer-PN expression. Its coefficient multiplying $\vv$ is supplied separately as
\texttt{ScalarDipoleCMVelocityCoefficient}; it includes $c^{-3}$ and
is not part of an integer-PN coefficient.

\subsection{Center-of-mass flux}\label{app:flux-coefficients}
The notebook supplies the coefficients completing the integer-PN polynomials in Eqs.~\eqref{eq:flux-CM-T} and \eqref{eq:flux-CM-S}. A subscript $n;ijk$ records the relative PN order and the powers of $Gm/r$, $v^2$, and $\dot r^2$, respectively. Coefficients absent within the stated PN orders and polynomial degrees are zero. Body-exchange symmetry is understood with $\Delta\to-\Delta$.
The quantities \texttt{FT[n,i,j,k]} and \texttt{FS[n,i,j,k]}
represent $F^T_{n;ijk}$ and $F^S_{n;ijk}$, respectively. No mass,
length, or velocity prefactors are absorbed into these coefficients.
To reconstruct the instantaneous power, first use the polynomials in
Eqs.~\eqref{eq:flux-CM-T} and \eqref{eq:flux-CM-S}, then add the
half-integer corrections displayed in Sec.~\ref{sec:flux}.
The notebook gives the latter as \texttt{TensorFluxHalf} and
\texttt{ScalarFluxHalf}, including their full prefactors.
The tensor term includes the nonlinear propagation contribution
adopted from Ref.~\cite{BBT2022}; tail integrals are not included.

\subsection{Circular-orbit reduction}\label{app:circular}
The generic instantaneous flux is first obtained by differentiating the
moments and reducing their derivatives with the required equations of
motion. Only then do we impose the circular conditions. This order
retains acceleration and higher-derivative terms that would be lost by
setting the radial velocity to zero prematurely.

\paragraph{Kepler relation.}
For a conservative circular orbit the relative acceleration is
$\va=-r\Omega^2\vn$, with $\dot r=0$ and $v^2=r^2\Omega^2$.
Taking the radial component of the relative acceleration therefore gives
\begin{equation}
\Omega^2=-\frac{\vn\cdot(\va_A-\va_B)}{r}
\bigg|_{\dot r=0,\ v^2=r^2\Omega^2}.
\label{eq:circular-radial-start}
\end{equation}
The right-hand side itself depends on $v^2$.
We solve this equation iteratively, using the Newtonian circular velocity
in the 1PN terms first, and including its 1PN correction when solving
at 2PN. In terms of the dimensionless separation
$\gamma=G_{AB}m/(c^2r)$, the result is
\begin{equation}
\Omega^2=\frac{G_{AB}m}{r^3}
\left[1+K_1\gamma+K_2\gamma^2+\order(\gamma^3)\right].
\label{eq:Kepler}
\end{equation}
The coefficients use the linear harmonic coordinates of the source
calculation; $\gamma$ is unrelated to $\bar\gamma_{AB}$:
\begin{align}
K_1={}&-3+\nu-\bar\gamma_{AB}-(\bar\beta_A+\bar\beta_B)
+\Delta(\bar\beta_A-\bar\beta_B),\label{eq:Kepler-K1}\\
K_2={}&\nu^2+\frac{41}{4}\nu+(5\nu-1)\bar\gamma_{AB}
-\frac{2\nu+1}{4}\bar\gamma_{AB}^2
+\left(4+\bar\gamma_{AB}+\frac\nu2\right)(\bar\beta_A+\bar\beta_B)
-\Delta\left(4+\bar\gamma_{AB}-\frac{5\nu}{2}\right)(\bar\beta_A-\bar\beta_B)\nonumber\\
&+\left(\frac12-\nu\right)(\delta_A+\delta_B)
+\frac\Delta2(\delta_A-\delta_B)
+\left(\frac14-\frac\nu2\right)(\epsilon_A+\epsilon_B)
-\frac\Delta4(\epsilon_A-\epsilon_B)+3\nu\zeta_{AB}.
\label{eq:Kepler-K2}
\end{align}

In the limit of vanishing scalar responses, these reduce to
$K_1=-3+\nu$ and $K_2=\nu^2+41\nu/4$ in these coordinates.

\paragraph{From separation to frequency.}
The frequency parameter in Sec.~\ref{sec:flux} obeys
\begin{equation}
x=\left(\frac{G_{AB}m\Omega}{c^3}\right)^{2/3}
=\gamma\left(1+K_1\gamma+K_2\gamma^2\right)^{1/3}
=\gamma+\frac{K_1}{3}\gamma^2+
\left(\frac{K_2}{3}-\frac{K_1^2}{9}\right)\gamma^3+\order(\gamma^4).
\label{eq:frequency-series-forward}
\end{equation}
Substituting the ansatz $\gamma=x+a x^2+b x^3$ into this series
and setting the coefficients of $x^2$ and $x^3$ in the difference
to zero determines $a$ and $b$:
\begin{equation}
\gamma=x-\frac{K_1}{3}x^2+\frac{K_1^2-K_2}{3}x^3+\order(x^4).
\label{eq:frequency-conversion}
\end{equation}
Both this inversion and the correction to $v^2$ must be retained when
converting the flux to frequency.

\paragraph{Circular flux at fixed separation.}
In the generic flux expressions
$P_T^{\rm inst}$ and $P_S^{\rm inst}$ in
Eqs.~\eqref{eq:flux-CM-T} and \eqref{eq:flux-CM-S},
the index $k$ labels the power of $\dot r^2$ in each
monomial. Setting $\dot r=0$ therefore leaves only
the terms with $k=0$. In these remaining terms, we substitute
\begin{equation}
\frac{Gm}{r}=\frac{c^2\gamma}{1+\alpha_A\alpha_B},\qquad
v^2=c^2\gamma\left(1+K_1\gamma+K_2\gamma^2\right).
\label{eq:circular-monomial-substitution}
\end{equation}
To display the origin of the frequency-conversion terms, write
\begin{align}
P_{T,\rm circ}^{\rm inst}&=\frac{c^5\nu^2\gamma^5}{G(1+\alpha_A\alpha_B)^4}
\left(H_0^T+H_1^T\gamma+H_2^T\gamma^2\right),\nonumber\\
P_{S,\rm circ}^{\rm inst}&=\frac{c^5\nu^2\gamma^4}{G(1+\alpha_A\alpha_B)^4}
\left(H_0^S+H_1^S\gamma+H_2^S\gamma^2\right).
\label{eq:circular-H-polynomials}
\end{align}
Here $H_n^T$ and $H_n^S$ are temporary polynomial coefficients,
not additional source moments. For either channel, let $d_T=1$ and
$d_S=0$, and define
\begin{equation}
W_n^X=\sum_{i=0}^{n+d_X}
\frac{F^X_{n;i\,(n+d_X-i)\,0}}{(1+\alpha_A\alpha_B)^i},
\qquad X\in\{T,S\}.
\label{eq:circular-W}
\end{equation}
Expanding the velocity powers in Eq.~\eqref{eq:circular-monomial-substitution}
gives, using $F^T_{0;100}=0$,
\begin{align}
H_0^T&=F^T_{0;010},&
H_1^T&=W_1^T+K_1F^T_{0;010},&
H_2^T&=W_2^T+K_1\left[\frac{F^T_{1;110}}{1+\alpha_A\alpha_B}
+2F^T_{1;020}\right]+K_2F^T_{0;010},\nonumber\\
H_0^S&=F^S_{0;000},&
H_1^S&=W_1^S,&
H_2^S&=W_2^S+K_1F^S_{1;010}.
\label{eq:circular-H-coefficients}
\end{align}

\paragraph{Circular flux at fixed frequency.}
Let $p_T=5$ and $p_S=4$ denote the leading powers of $\gamma$.
Applying Eq.~\eqref{eq:frequency-conversion} to both the prefactor
$\gamma^{p_X}$ and the bracket in Eq.~\eqref{eq:circular-H-polynomials}
gives
\begin{align}
f_0^X&=\frac{H_0^X}{(1+\alpha_A\alpha_B)^2},\nonumber\\
f_1^X&=\frac{H_1^X-\frac{p_X}{3}K_1H_0^X}{(1+\alpha_A\alpha_B)^2},\nonumber\\
f_2^X&=\frac{H_2^X-\frac{p_X+1}{3}K_1H_1^X+
\left[\frac{p_X(p_X+5)}{18}K_1^2-\frac{p_X}{3}K_2\right]H_0^X}
{(1+\alpha_A\alpha_B)^2}.
\label{eq:circular-universal-conversion}
\end{align}
The leading coefficients are $f_0^T=32/5$ and
$f_0^S=(\alpha_A-\alpha_B)^2/3$.
Substituting the intermediate coefficients yields the formulas used
in Eqs.~\eqref{eq:ST-circular-T} and \eqref{eq:ST-circular-S}:
\begin{align}
f^T_1={}&\frac{1}{(1+\alpha_A\alpha_B)^2}
\left[\frac{F^T_{1;200}}{(1+\alpha_A\alpha_B)^2}
+\frac{F^T_{1;110}}{1+\alpha_A\alpha_B}+F^T_{1;020}-\frac23K_1F^T_{0;010}\right],\label{eq:circ-T1}\\
f^T_2={}&\frac{1}{(1+\alpha_A\alpha_B)^2}
\left[\sum_{i=0}^{3}\frac{F^T_{2;i\,(3-i)\,0}}{(1+\alpha_A\alpha_B)^i}
-K_1\left(\frac{2F^T_{1;200}}{(1+\alpha_A\alpha_B)^2}+\frac{F^T_{1;110}}{1+\alpha_A\alpha_B}\right)
+\left(\frac79K_1^2-\frac23K_2\right)F^T_{0;010}\right],\label{eq:circ-T2}\\
f^S_1={}&\frac{1}{(1+\alpha_A\alpha_B)^2}
\left[\frac{F^S_{1;100}}{1+\alpha_A\alpha_B}+F^S_{1;010}-\frac43K_1F^S_{0;000}\right],\label{eq:circ-S1}\\
f^S_2={}&\frac{1}{(1+\alpha_A\alpha_B)^2}
\left[\sum_{i=0}^{2}\frac{F^S_{2;i\,(2-i)\,0}}{(1+\alpha_A\alpha_B)^i}
+K_1\left(-\frac{5F^S_{1;100}}{3(1+\alpha_A\alpha_B)}-\frac23F^S_{1;010}\right)
+\left(2K_1^2-\frac43K_2\right)F^S_{0;000}\right].\label{eq:circ-S2}
\end{align}

The terms proportional to $K_1$ and $K_2$ thus have two origins:
the correction to the circular velocity and the conversion from
coordinate separation to orbital frequency. All coefficients on the
right-hand sides are supplied in the ancillary notebook, which also
provides the assembled $f_n^T$ and $f_n^S$ as \texttt{fT[n]} and
\texttt{fS[n]}.
The normalization does not divide by the leading scalar dipole charge,
so the equal-charge limit is regular whenever $G_{AB}\ne0$.

Finally, the half-integer corrections displayed in Sec.~\ref{sec:flux}
are proportional to $\dot r$ after the generic differentiation has
been completed. They vanish on the conservative circular orbit.
For an adiabatic quasi-circular inspiral, inserting the leading
radiation-driven radial drift into these terms adds another 1.5PN
suppression and places their effect beyond the orders retained here.
The circular formulas above therefore use the integer-PN coefficients,
while still representing only the instantaneous part of the power:
the omitted tail contributions do not in general vanish for circular
motion.

\twocolumngrid
\bibliographystyle{apsrev4-2}
\bibliography{references}
\end{document}